\pdfoutput=1

\documentclass{sig-alternate} 
\usepackage{mathptmx} 

\newcommand{\ignore}[1]{}
\usepackage{fancyhdr}
\usepackage[normalem]{ulem}
\usepackage[hyphens]{url}
\usepackage[sort,nocompress]{cite}
\usepackage[final]{microtype}
\usepackage{flushend}

\usepackage[bookmarks=true,breaklinks=true,letterpaper=true,colorlinks,linkcolor=black,citecolor=blue,urlcolor=black]{hyperref}

\usepackage[usenames,dvipsnames]{color}
\usepackage{xspace}
\usepackage{enumitem}
\usepackage{booktabs}
\usepackage{multirow}
\usepackage{listings}
\usepackage{setspace}
\usepackage{float}

\newcommand{\paratitle}[1]{\vspace{4pt}\noindent\textbf{#1}}
\newcommand{\incircle}[1]{\raisebox{.5pt}{\textcircled{\small #1}}}

\newcommand{\footnoteref}[1]{\textsuperscript{\ref{#1}}}
\newcommand{\changes}[0]{}

\newcommand{\cpuwriteset}{\texttt{CPUWriteSet}\xspace}
\newcommand{\pimreadset}{\texttt{PIMReadSet}\xspace}
\newcommand{\pimwriteset}{\texttt{PIMWriteSet}\xspace}

\fancypagestyle{firstpage}{
  \fancyhf{}
\setlength{\headheight}{50pt}

  \pagenumbering{arabic}
}

\title{LazyPIM: Efficient Support for Cache Coherence \\in Processing-in-Memory Architectures\vspace{-40pt}}

\usepackage[auth-lg, affil-it]{authblk}
\makeatletter
\renewcommand\AB@affilsepx{}
\makeatother

\author[${*}$]{Amirali Boroumand}
\author[${*}$]{Saugata Ghose}
\author[${*}$]{Minesh Patel$^{\dagger}$}
\author[${*}$]{Hasan Hassan$^{\dagger}$}
\author[${*}$]{Brandon Lucia}
\author[${\star}{*}$]{\\Nastaran Hajinazar} 
\author[${*}$]{Kevin Hsieh}
\author[${\Phi}$]{Krishna T. Malladi}
\author[${\Phi}$]{Hongzhong Zheng}
\author[$\dagger{*}$]{Onur Mutlu}

\affil[${*}$]{Carnegie Mellon University\quad\quad} 
\affil[${\dagger}$]{ETH Z{\"u}rich\quad\quad}
\affil[${\ddagger}$]{TOBB ET{\"U}\protect\\\vspace{-3pt}}
\affil[${\star}$]{Simon Fraser University\quad\quad}
\affil[${\Phi}$]{Samsung Semiconductor, Inc.\vspace{-10pt}}

\begin{document}
\maketitle
\thispagestyle{firstpage}
\pagestyle{plain}

\tolerance 1414
\hbadness 1414
\emergencystretch 1.5em
\hfuzz 0.3pt
\clubpenalty=10000
\widowpenalty=10000
\vfuzz \hfuzz
\raggedbottom

\setstretch{0.941}
\fontdimen2\font=0.38ex

\begin{abstract}

\emph{Processing-in-memory} (PIM) architectures have seen an increase in
popularity recently, as the high internal bandwidth available within 3D-stacked 
memory provides greater incentive to move some computation into the logic
layer of the memory.  To maintain program correctness, the portions of a
program that are executed in memory must remain coherent with the portions of
the program that continue to execute within the processor.  Unfortunately,
PIM architectures cannot use traditional approaches to cache coherence
due to the high off-chip traffic consumed by coherence messages, which, as
we illustrate in this work, can 
undo the benefits of PIM execution for many data-intensive applications.

We propose \emph{LazyPIM}, a new hardware cache coherence
mechanism designed specifically for PIM.  
Prior approaches for coherence in PIM are ill-suited to applications that share
a large amount of data between the processor and the PIM logic.
LazyPIM uses a combination of
speculative cache coherence and compressed coherence signatures to greatly
reduce the overhead of keeping PIM coherent with the processor,
even when a large amount of sharing exists. 
We find that LazyPIM improves average performance across a range of 
data-intensive PIM applications by 19.6\%, reduces off-chip traffic by 30.9\%,
and reduces energy consumption by 18.0\%, over the best prior approaches to PIM coherence. 

\end{abstract}



\section{Introduction}
\label{sec:intro}

For many modern data-intensive applications, main memory bandwidth 
continues to be a limiting factor, as data movement contributes
to a large portion of a data-intensive application's execution time
and energy consumption. In fact, modern computing platforms are
unable to scale performance linearly with the amount of data processed
per each computing node~\cite{Ahn:2015:SPA:2749469.2750386}. The emergence of memory-intensive big-data
applications, such as in-memory databases and graph processing frameworks,
exacerbates this problem. These workloads often exhibit frequent 
data movement and random memory access patterns, placing high pressure
on the memory system and making memory bandwidth a first-class problem
for those workloads.

\changes{Recent advances in 3D-stacked technology enable promising solutions
to alleviate the memory bandwidth problem.
To exploit the high internal bandwidth available within 3D-stacked DRAM, }
several recent works explore \emph{processing-in-memory} (PIM),
also known as \emph{near-data processing} (e.g.,~\cite{Ahn:2015:SPA:2749469.2750386, 
Ahn:2015:PIL:2749469.2750385, Mingyu:PACT, guo2014wondp, toppim, seshadri2013rowclone, Seshadri:2015:ANDOR, Hsieh:2016:ISCA:TOM, pugsley2014ndc, JAFAR, 7056040, xu2015scaling}),
where the processor dispatches parts of the application (\emph{PIM kernels}) for execution at compute units (\emph{PIM cores}) within 
DRAM. The specific functionality enabled by these
PIM cores differs from proposal to proposal, with some work implementing
specialized accelerators~\cite{Ahn:2015:SPA:2749469.2750386,PICA}
while others add simple general-purpose cores~\cite{Mingyu:PACT,Ahn:2015:PIL:2749469.2750385} or GPU cores~\cite{toppim} in the logic layer.


%

To ensure correct execution, PIM architectures require coordination between the
processor cores and the PIM cores.  One of the primary challenges for 
coordination is cache coherence, which ensures that cores always use the
correct version of the data (as opposed to stale data that does not include
updates that were performed by other cores).  If the PIM cores are
coherent with the processor cores, the PIM programming model becomes
relatively simple, as it then becomes similar to conventional multithreaded
programming.  Employing a common programming model for PIM that most
programmers are familiar with can facilitate the widespread adoption of PIM.



However, it is impractical for PIM to utilize a
traditional cache coherence protocol (e.g., MESI), as this forces a large number of coherence messages to
traverse the narrow off-chip bus that exists between PIM cores and the processor cores, 
potentially undoing the benefits of high-bandwidth PIM
execution. Most prior works on PIM assume that only a limited amount of data sharing 
occurs between the PIM kernels and the processor threads of an application.
Thus, they sidestep coherence by employing solutions that
restrict PIM to execute on non-cacheable data
(e.g.,~\cite{Ahn:2015:SPA:2749469.2750386, 7056040, toppim}) or force the processor cores to
flush or not access any data that could \emph{potentially} be used by
PIM (e.g.,~\cite{7056040, Mingyu:PACT, Seshadri:2015:ANDOR, seshadri2013rowclone, guo2014wondp,
Ahn:2015:PIL:2749469.2750385}).


We comprehensively study two important classes of data-intensive applications, graph processing
frameworks and in-memory databases, where we find there is strong potential
for improvement using PIM.
We make two \emph{key observations} based on our analysis of these data-intensive
applications: 
(1)~some portions of the applications are
better suited for execution in processor threads, and these portions often concurrently access the same region
of data as the PIM kernels, leading to \emph{significant data sharing}; and 
(2)~poor handling of coherence eliminates a significant portion of the performance \changes{and energy} benefits of PIM.
As a result, we find that a good coherence mechanism is \emph{required} to retain
the full benefits of PIM across a wide range of applications (see Section~\ref{sec:motivation}) while
maintaining the correct execution of the program. \textbf{Our goal} in this work
is to propose a cache coherence mechanism for PIM architectures that
\emph{logically behaves} like traditional coherence, but retains all of the benefits of PIM.

To this end, we propose \emph{LazyPIM}, a new cache coherence mechanism that
efficiently batches coherence messages sent by the PIM cores.  During PIM kernel execution, a PIM core 
\emph{speculatively} assumes that it has acquired coherence permissions without 
sending a coherence message, and maintains all data updates speculatively in its 
cache.  Only when the kernel finishes execution, the processor receives compressed information from
the PIM core, and checks if any coherence conflicts occurred. 
LazyPIM uses compressed \emph{signatures}~\cite{Bloom:1970:STH:362686.362692}
to track potential conflict information efficiently. 
Under the LazyPIM execution model, a conflict occurs only when the PIM core 
reads data that the processor updated (i.e., wrote to) before the conflict 
check takes place (see Section~\ref{sec:mech:conflicts}).
If a conflict exists, 
the dirty cache lines in the processor 
are flushed, and the PIM core rolls back and re-executes the kernel. 
Our execution model \emph{for the PIM cores}
is similar to \emph{chunk-based execution}~\cite{bulksc} (i.e., each 
\emph{batch} of consecutive instructions executes atomically), which prior 
work has harnessed for various purposes such as transactional memory~\cite{tcc} and 
deterministic execution~\cite{dmp}.
Unlike past works, the processor in LazyPIM executes conventionally and \emph{never rolls back},
which can make it easier to adopt PIM in general-purpose systems.

We find that LazyPIM is highly effective at providing cache coherence for PIM
architectures, and avoids the high overheads of previous solutions even when 
a high degree of data sharing exists between the PIM kernels and the processor
threads.  Due to the high overhead of prior approaches to PIM coherence, PIM
execution often fares worse than if we executed the workloads entirely on
the CPU, with PIM execution providing worse performance, consuming more 
off-chip traffic, and consuming more energy in a number of cases.  
\changes{In contrast, for \emph{all} of our workloads, LazyPIM
retains most of the benefits of an ideal PIM mechanism that has no penalty for
coherence, coming
within 9.8\% and 4.4\% of the average performance and energy, respectively, for our
16-thread data-intensive workloads.
LazyPIM reduces the execution time and energy
by 66.0\% and 43.7\%, respectively, over CPU-only execution.}

We make the following key contributions in this work:
\vspace{-5pt}
\begin{itemize}[leftmargin=1em]
  \itemsep 0pt
  \item We demonstrate that previously-proposed coherence mechanisms for PIM
    are not a good fit for workloads
    where a large amount of data sharing occurs between
    the PIM kernels and the processor threads.  Prior mechanisms either 
    take overly conservative approaches to sharing or 
    generate high off-chip traffic, oftentimes undoing the benefits that PIM architectures provide.
  \item We propose LazyPIM, a new hardware coherence mechanism for PIM.  Our approach
    (1)~reduces the off-chip traffic between the PIM 
    cores and the processor, (2)~avoids the costly overheads of prior 
    approaches to provide coherence for PIM, and (3)~retains the same 
    logical coherence behavior as architectures without PIM to keep programming 
    simple.
  \item We show that on average for our 16-thread applications, LazyPIM improves 
    performance by 19.6\%,
    reduces off-chip traffic by 30.9\%, and
    reduces energy consumption by 18.0\% over the best prior coherence approaches, respectively, for each metric.
    LazyPIM enables PIM execution to always outperform CPU-only execution.
\end{itemize}



\section{Background}
\label{sec:background}
\label{sec:bkgd}


In this section, we provide necessary background on PIM architectures.
First, we study how 3D-stacked memory can deliver high internal bandwidth (Section~\ref{sec:bkgd:3d}).  
Then, we explore prior works on PIM
(Section~\ref{sec:bkgd:pim}).  
Finally, we discuss the baseline PIM architecture that we study for this paper
(Section~\ref{sec:bkgd:baseline}).

\subsection{3D-Stacked Memory}
\label{sec:bkgd:3d}

In recent years, a number of new DRAM architectures have been proposed to
take advantage of 3D circuit integration technology~\cite{loh2008stacked, lee2016smla}.  These architectures stack
multiple layers of DRAM arrays together within a single chip. 3D-stacked memory employs \emph{through-silicon vias} 
(TSVs), which are vertical wires that connect all stack layers together.
Due to the available density of TSVs, 3D-stacked memories are able to
provide much greater bandwidth between stack layers than they can provide
off-chip.
Several 3D-stacked memory architectures,
such as High Bandwidth Memory (HBM)~\cite{hbm}
and the Hybrid Memory Cube (HMC)~\cite{hmcspec2},
also provide the
ability to perform low-cost logic integration directly within memory.  Such
architectures dedicate a \emph{logic layer} within the stack 
for logic circuits.  Within the logic layer, architects can 
implement functionality that interacts with both the DRAM cells in the other
layers (using the TSVs) and the processor. 
For example, HMC uses part of the logic layer to buffer memory requests from
the processor and perform memory scheduling~\cite{hmcspec2}.


\subsection{Processing-in-Memory}
\label{sec:pim-background}
\label{sec:bkgd:pim}

\changes{The origins of PIM go back to proposals from the 1970s, where small processing elements
were combined with small amounts of RAM to provide a distributed array of
memories that perform computation~\cite{shaw1981non, stone1970logic}.
Early works such as EXECUBE~\cite{kogge1994execube} and IRAM~\cite{patterson1997case}
 added logic within DRAM
 to perform vector operations. Later works~\cite{kang2012flexram,Draper:2002:ADP:514191.514197,Mai:2000:SMM:339647.339673,oskin1998active} proposed more versatile substrates that increased the flexibility and compute power available within DRAM.
These proposals had limited to no adoption, as the proposed logic integration was too costly and did not solve many 
of the obstacles facing PIM.}

With the advent of 3D-stacked memories, we have seen a resurgence of PIM
 proposals.
Recent PIM proposals (e.g.,~\cite{Ahn:2015:SPA:2749469.2750386, 7056040, Ahn:2015:PIL:2749469.2750385, Mingyu:PACT, toppim, guo2014wondp, Hsieh:2016:ISCA:TOM, PICA})
add compute units within the logic layer to exploit the high bandwidth available.
These works have primarily focused on the design of the underlying logic that is
placed within memory, and in many cases propose special-purpose PIM architectures
that cater only to a limited set of applications.

\subsection{Baseline PIM Architecture}
\label{sec:baseline-organization}
\label{sec:bkgd:baseline}

\begin{figure}[h]
    \vspace{-5pt}
    \centering
        \includegraphics[width=0.73\linewidth]{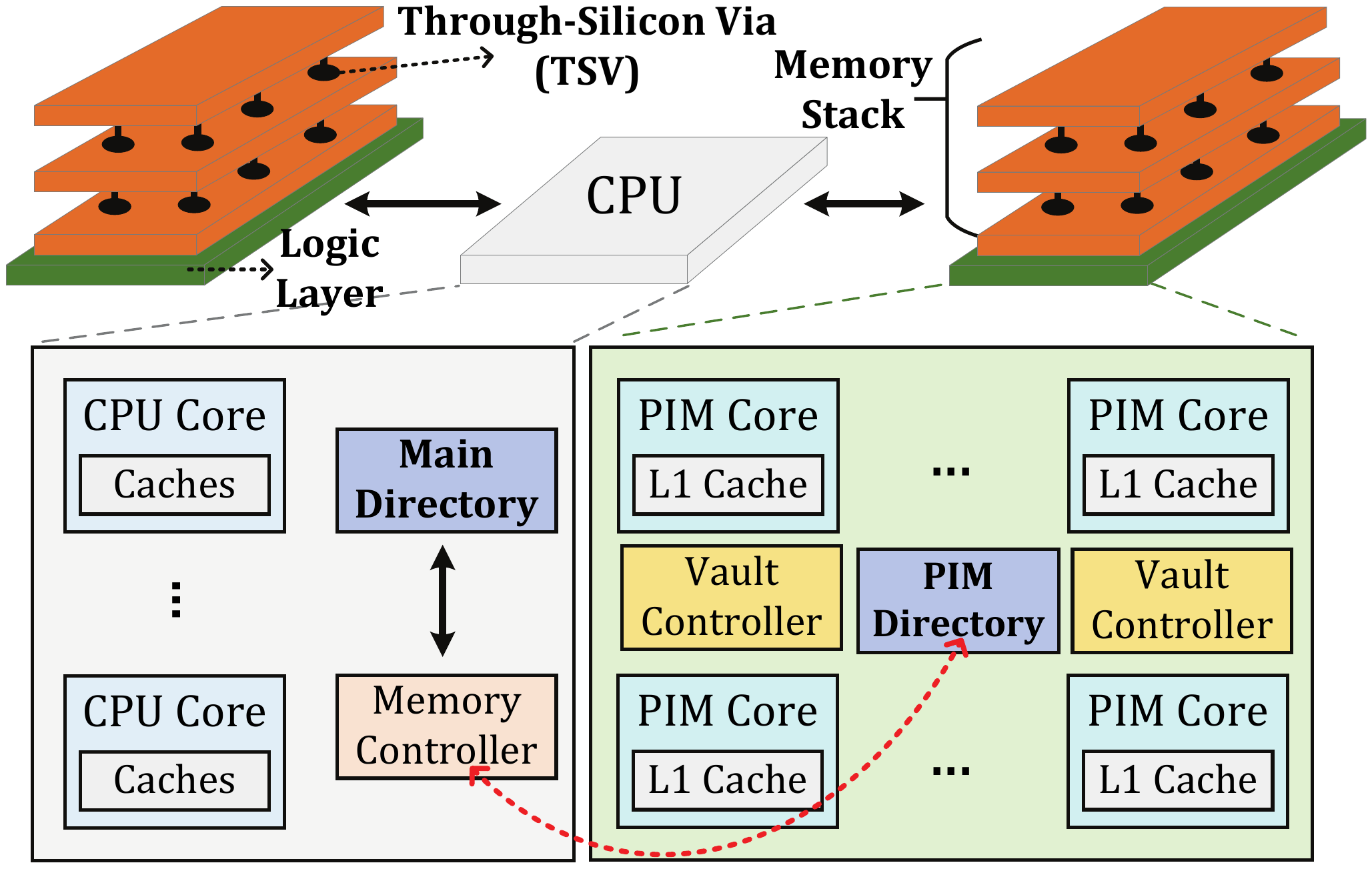}%
    \vspace{-10pt}%
    \caption{High-level organization of PIM architecture.}
    \label{fig:baseline}
  \vspace{-5pt}
\end{figure}

Figure~\ref{fig:baseline} shows the baseline organization of the PIM
architecture in our paper.
We refer to the compute units within the main processor as \emph{processor
cores}, which execute \emph{processor threads}.  We refer to the
compute units within memory as \emph{PIM cores}, which execute \emph{PIM
kernels} that are invoked by the processor threads.
\changes{Note that LazyPIM can be used with any type of compute units that contain memory.}
In our evaluation, we assume that the compute units 
within memory consist of simple 
\emph{in-order} cores.  These PIM cores, which are ISA-compatible with the 
processor cores, are much weaker in terms of performance, as
they lack large caches and sophisticated ILP techniques, 
and frequently do not implement all of the ISA. 
These weaker cores are more practical to implement within the DRAM logic layer.

Each PIM core includes small private L1 I/D caches. The data cache is kept coherent using
a local MESI directory called the \emph{PIM coherence directory}, 
which resides within the DRAM logic layer.
The processor caches maintain coherence using the traditional on-chip directory.
The processor directory acts as the main coherence point for the system, interfacing 
with both the processor caches and the PIM coherence directory.
As prior PIM works~\cite{Ahn:2015:SPA:2749469.2750386, 7056040} have done,
we assume that direct segments~\cite{basu2013efficient}
are used for PIM data, and that PIM kernels operate only on physical addresses.


\section{Motivation}
\label{sec:motivation}

Systems with multiple independent processor cores rely on cache coherence to seamlessly
allow multiple threads to exchange data with each other in the presence of a
cache.  If we treat the PIM cores as additional independent cores that remain
coherent with the processor cores, we can greatly simplify the PIM programming
model, by making it behave similarly to the behavior programmers expect for
multithreaded programming.  While recent PIM proposals acknowledge the need for
some form of coherence between the PIM cores and processor cores~\cite{JAFAR,7056040,Mingyu:PACT},
these works have largely sidestepped the issue by 
assuming that there is only a limited amount of data sharing that takes places 
between PIM kernels and processor threads during PIM kernel execution. \changes{In
this section, we show that this assumption does not hold for many 
classes of applications that are well-suited for PIM execution.}

\subsection{Applications with High Data Sharing}
\label{sec:motivation:sharing}

An application benefits the most from PIM execution when its memory-intensive parts, 
which often exhibit poor locality and contribute to a large portion of 
the overall execution time, are dispatched to the PIM cores.
Executing on the PIM cores allows the
memory-intensive parts to benefit the most from the high bandwidth and 
low latency access available at the logic layer of the 3D-stacked DRAM. 
On the other hand, the compute-intensive parts of an application,
typically the portions of code that exhibit high locality, \emph{must remain 
on the processor cores} to maximize performance~\cite{Ahn:2015:PIL:2749469.2750385, Hsieh:2016:ISCA:TOM, Mingyu:PACT}.
We discuss how we find the appropriate partitions between memory-intensive
and compute-intensive parts of our applications in 
Section~\ref{sec:methodology:kernels}.

Prior work mostly assumes that there is only a limited amount of data sharing between the PIM kernels and the processor threads. 
However, \emph{this is not the case} for many important applications, such as graph and database workloads. 
One class of examples is multithreaded graph processing frameworks,
such as Ligra~\cite{ligra}, where multiple threads operate on the same shared in-memory graph~\cite{Seraph,ligra}. Each thread executes a
graph algorithm, such as PageRank. We studied a number of these algorithms~\cite{ligra} and found that when we convert
each of them for PIM execution, only some portions of each algorithm were well-suited for PIM, while the remaining portions
performed better if they stayed on the processor to exploit locality using large caches. This observation was also made by prior work~\cite{Ahn:2015:PIL:2749469.2750385}. With this partitioning,
some threads execute on the processor cores while other threads (sometimes concurrently) execute on the PIM cores, with all of the
threads sharing the graph and other intermediate data structures.


Another example is a modern in-memory database that supports hybrid transactional/analytical processing (HTAP) workloads.
Today, analytic and transactional behaviors are being combined in a single hybrid database system, as observed in several
academic and industrial databases~\cite{SAP,stonebraker2013voltdb,MemSQL}, thanks to the need to perform real-time analytics on transactional data. The
analytic portions of these hybrid databases are well-suited for PIM execution, as analytical queries have long lifetimes and touch
a large number of database rows, leading to a large amount of random memory accesses and data movement~\cite{kocberber2013meet, Hash:NME, JAFAR}. On
the other hand, even though transactional queries access the same data, they likely perform better if they stay on the processor,
as they are short-lived and are often latency sensitive, accessing only a few rows each, which can easily fit in processor caches. In such workloads, concurrent accesses
from both PIM kernels and processor threads are inevitable, as analytical queries benefit from executing near the data while
transactions perform best on processor cores.












\subsection{Prior Approaches to PIM Coherence}
\label{sec:motivation:naive}

Shared data needs to remain
 coherent between the processor and PIM cores.
In an ideal scenario, there would be no communication overhead
for maintaining coherence.
Traditional, or \emph{fine-grained}, coherence protocols (e.g., MESI~\cite{Papamarcos:1984:LCS:800015.808204,Goodman:1983:UCM:800046.801647})
have several qualities well suited for pointer-intensive
data structures, such as those in graph workloads and databases.
The path taken while traversing pointers during pointer chasing is not known ahead of time. As a result,
 even though a thread often accesses only a few dispersed pieces of the data structure,
 a coarse-grained mechanism has no choice but to acquire coherence
permissions for the entire data structure.
Fine-grained coherence allows the processor or PIM to acquire permissions for only
the pieces of data within the shared structure that are actually accessed.
In addition, fine-grained coherence 
can ease programmer effort when developing PIM applications, as multithreaded programs already use this programming model.

\begin{figure}[h]
    \vspace{-11pt}
    \centering
        \centering
        \includegraphics[width=\linewidth]{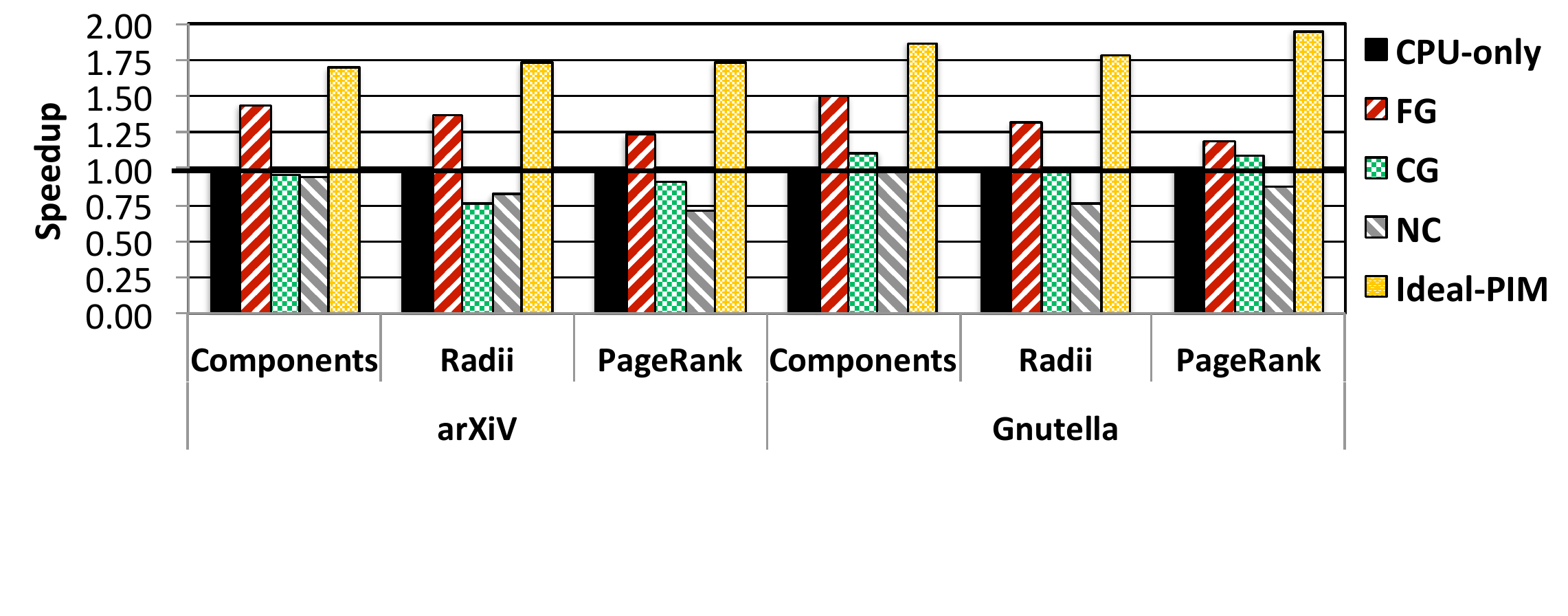}%
    \vspace{-35pt}%
    \caption{PIM speedup with 16 threads, normalized to CPU-only execution.{\protect\footnoteref{foot:meth}}}
    \label{fig:motiv}
  \vspace{-6pt}
\end{figure}

Unfortunately, if a PIM core participates in traditional coherence, it would have to send
 a message for \emph{every cache miss} to the main directory in the processor, over a narrow pin-limited bus (we call this \emph{PIM
 coherence traffic}). 
Figure~\ref{fig:motiv} shows the speedup of PIM with different coherence
mechanisms for a few example graph workloads, normalized to a \emph{CPU-only}
baseline (where the whole application runs on the
processor).\footnote{\label{foot:meth}See Section~\ref{sec:methodology} for our
methodology.} To illustrate the impact of inefficient mechanisms, we also show
the performance of an \emph{ideal} mechanism where there is no performance
penalty for coherence (\emph{Ideal-PIM}).  As shown in Figure~\ref{fig:motiv},
PIM with fine-grained coherence (\emph{FG}) \changes{eliminates a significant
portion of the Ideal-PIM improvement and often performs only slightly better
than CPU-only execution due to its high PIM coherence traffic.} As a result,
while Ideal-PIM can improve performance by an average of \changes{84.2\%}, FG
only attains performance improvements of \changes{38.7\%}.

To reduce the impact of PIM coherence traffic, there are three general alternatives to fine-grained
coherence that have been proposed for PIM execution: (1)~coarse-grained coherence, (2)~coarse-grained locks,
 and (3)~making PIM data non-cacheable in the processor. 
 
\paratitle{Coarse-Grained Coherence.}
One approach to reduce PIM coherence traffic is to 
apply \emph{coarse-grained coherence}.  For example, we can 
maintain a single
coherence entry for \emph{all} of the PIM data.  Unfortunately, this can 
still incur high overheads, as the processor
must flush \emph{all} of the dirty cache lines within the PIM data 
region \emph{every time} the PIM core acquires permissions, 
\emph{even if the PIM kernel may not access most of the data}.
For example, with just four processor threads, the number of cache lines flushed for PageRank is 
227x the number of lines \emph{actually required by the PIM kernel}.\footnoteref{foot:meth}
Coherence at a smaller granularity, such as maintaining a coherence entry for
each page~\cite{Mingyu:PACT}, does not cause flushes for pages that are not
accessed by the PIM kernel.  However, this optimization is not beneficial for many data-intensive applications that perform
\emph{pointer chasing}, where a large number of pages are
\emph{accessed} non-sequentially, but only a \emph{few lines} in each page are used, forcing
the processor to flush \emph{every} dirty page.



\paratitle{Coarse-Grained Locks.}
Another drawback of coarse-grained coherence is that data can ping-pong
between the processor and the PIM cores whenever the PIM data region is concurrently
accessed by both.  \emph{Coarse-grained
locks} avoid ping-ponging by having the PIM cores acquire \emph{exclusive} access to a region for
the duration of the PIM kernel.  However, 
coarse-grained locks greatly restrict performance. 
Our application study shows that PIM kernels and processor threads often
work in parallel on the same data region, and coarse-grained locks frequently
cause thread serialization. 
For our graph applications running on
a representative input (Gnutella), coarse-grained locks block an
average of 87.9\% of the processor cores' memory accesses during PIM kernel
execution. 
The frequent blocking, along with the large number of unnecessary flushes, results in
significant performance loss, as shown in Figure~\ref{fig:motiv}.
 PIM with coarse-grained locks (\emph{CG} in Figure~\ref{fig:motiv})
 can in some cases eliminate the entire benefit of Ideal-PIM execution, 
and performs 
\changes{1.4\%} \emph{worse} than CPU-only execution, on average.
We conclude that using coarse-grained locks is not suitable for many
important applications for PIM execution.

\paratitle{Non-Cacheable PIM Data.}
Another approach sidesteps coherence, by marking the PIM data region 
as \emph{non-cacheable} in the processor~\cite{Ahn:2015:SPA:2749469.2750386}, 
so that DRAM always contains up-to-date data.  For applications where PIM data
is almost exclusively accessed by the PIM cores, this incurs little
penalty, but for many applications, the processor also accesses PIM data
often.  For our graph applications with a representative input (arXiV),\footnoteref{foot:meth}
the processor cores generate 38.6\% of the total number of accesses to PIM data.
With so many processor accesses,
making PIM data non-cacheable results in high performance and 
bandwidth overhead. As shown in Figure~\ref{fig:motiv}, 
marking PIM data as non-cacheable (\emph{NC}) fails to retain any of the benefits of Ideal-PIM,
and \emph{always} performs worse 
than CPU-only execution (\changes{3.2\%} worse on average).
 Therefore, while this approach avoids the overhead of
coarse-grained mechanisms, it is a poor fit for applications
that rely on processor involvement, and thus restricts when PIM is effective.




We conclude that prior approaches to PIM coherence eliminate a significant
portion of the benefits of PIM when data sharing occurs, due to their high 
coherence overheads. In fact, they sometimes cause PIM execution to consistently 
degrade performance. Thus, an \emph{efficient} alternative to fine-grained 
coherence is necessary to retain PIM benefits across a wide range
of applications, including those applications where the overhead of coherence
made them a poor fit for PIM in the past. 



\section{LazyPIM Coherence Behavior}
\label{sec:mech}



\begin{figure}[h]
\centering
\vspace{-5pt}
\includegraphics[width=0.75\linewidth]{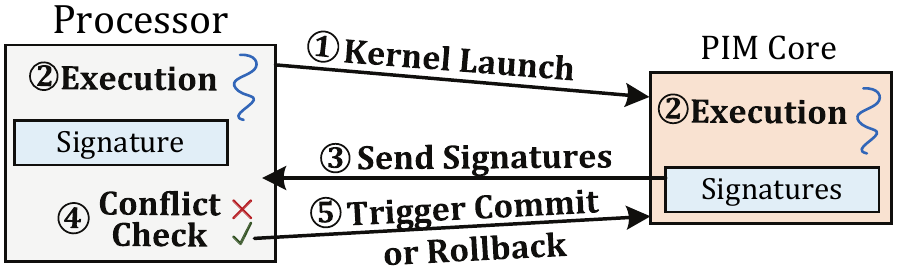}%
\vspace{-10pt}%
\caption{High-level operation of LazyPIM.}
\label{fig:operation}
\vspace{-5pt}
\end{figure}

Our goal is to design a coherence mechanism that maintains the logical behavior
of traditional coherence while retaining the large performance \changes{and energy} benefits of PIM.
To this end, we propose 
\emph{LazyPIM}, a new coherence mechanism that lets PIM kernels
\emph{speculatively} assume that they have the required permissions from the
coherence protocol, \emph{without} actually sending off-chip messages to the main
(processor) coherence directory during execution.
Figure~\ref{fig:operation} shows the high-level operation of LazyPIM.
The processor launches a kernel on a PIM core (\incircle{1} in 
Figure~\ref{fig:operation}), allowing the PIM kernel to execute with 
speculative coherence permissions while the processor thread executes
concurrently (\incircle{2}).  Coherence states are updated only
\emph{after} the PIM kernel completes, at which point the PIM core transmits a
single batched coherence message (i.e., compressed \emph{signatures} containing 
\emph{all} addresses that the PIM kernel read from or wrote to) back to the 
processor coherence directory (\incircle{3}). The directory checks to see 
whether any \emph{conflicts} occurred (\incircle{4}). If a conflict exists, the PIM kernel
\emph{rolls back} its changes, conflicting cache lines are written back by the processor to 
DRAM, and the kernel re-executes.  If no conflicts exist, speculative data within 
the PIM core is \emph{committed}, and the processor coherence directory is
updated to reflect the data held by the PIM core (\incircle{5}).  Note that in LazyPIM, the
processor \emph{always} executes \emph{non-speculatively}, which ensures
minimal changes to the processor design, thereby enabling easier adoption of 
PIM.

LazyPIM avoids the pitfalls of the mechanisms discussed in
Section~\ref{sec:motivation}.  With its compressed signatures, LazyPIM causes much
less PIM coherence traffic than traditional coherence.  Unlike coarse-grained
coherence and coarse-grained locks, LazyPIM checks coherence \emph{only after} 
it completes PIM execution, avoiding the need to
unnecessarily flush a large amount of data.  LazyPIM also allows for efficient concurrent
execution of processor threads and PIM kernels: by executing speculatively,
the PIM cores do \emph{not} invoke coherence requests during concurrent execution,
avoiding data ping-ponging and allowing processor threads to continue using
their caches.


In the remainder of this section, we study three key aspects of coherence 
behavior under LazyPIM.  
Section~\ref{sec:mech:conflicts} describes the conflicts that can occur when
PIM kernels execute speculatively.
Section~\ref{sec:mech:partial} discusses how LazyPIM uses partial PIM kernel
commits to reduce the overhead of rollback when a conflict occurs.
Section~\ref{sec:mech:multicore} describes how multiple PIM kernels interact with
each other without violating program correctness.
\changes{Section~\ref{sec:mech:synch} discusses how LazyPIM supports various synchronization primitives.}
We discuss the architectural support required to implement LazyPIM in
Section~\ref{sec:arch}.

\subsection{Conflicts} 
\label{sec:mech:conflicts}

In LazyPIM, a PIM kernel \emph{speculatively} assumes during execution that it
has coherence permissions on a cache line, without checking the processor
coherence directory.  In the meantime, the processor continues to execute
\emph{non-speculatively}.  To resolve speculation without violating the memory consistency
model, LazyPIM
provides \emph{coarse-grained atomicity}, where all PIM memory updates are
treated as if they \emph{all} occur \emph{at the moment that a PIM kernel
commits}.
At this point, a \emph{conflict} may
be detected on a cache line read by the PIM kernel.

We explore three possible interleavings of read and write operations to a cache
line where the use of coarse-grained atomicity by LazyPIM could result in
unintended memory orderings.
In this discussion, we assume a sequentially consistent memory model, though 
this reasoning can easily be applied to other memory consistency models, such 
as x86-TSO.

\paratitle{PIM Read and Processor Write to the Same Cache Line.}
\textit{This is a conflict.}
Any PIM read during the kernel execution should be
ordered \emph{after} the processor write (RAW, or read-after-write). 
However, because the PIM core does not check coherence while it is in the middle of
kernel execution, the PIM read operation reads potentially stale data from
 DRAM instead of the output of the processor write (which
sits in the processor cache). This is detected by the LazyPIM architecture,
resulting in rollback and re-execution of the PIM kernel
after the processor write is flushed to DRAM.

\paratitle{Processor Read and PIM Write to the Same Cache Line.} 
\textit{This is not a conflict.}
With coarse-grained atomicity, any read by the processor during PIM execution 
is ordered \emph{before} the PIM core's write (WAR, or write-after-read).
As a result, the processor should \emph{not} read the value written by PIM. 
LazyPIM ensures this by marking the PIM write as speculative and maintaining 
the data in the PIM cache, preventing processor read requests from 
seeing any data flagged as speculative.

Note that if the programmer wants the processor's read to see the PIM core 
write, they need to use a synchronization primitive (e.g., memory 
barrier) to enforce ordering between the processor and the PIM core. This 
scenario is identical to conventional multithreading, where
an explicit ordering of memory operations must be enforced 
using synchronization primitives, which are supported by LazyPIM \changes{(see Section~\ref{sec:mech:synch})}.


\paratitle{Processor Write and PIM Write to the Same Cache Line.}
\textit{This is not a conflict}. \changes{With coarse-grained atomicity, any
write by the processor during PIM kernel execution is ordered before the PIM
core's write (WAW, or write-after-write) since the PIM core write effectively
takes place \emph{after the PIM kernel finishes}. When the two writes modify
different words in the same cache line, LazyPIM uses a per-word dirty bit mask
in the PIM L1 cache to merge the writes, similar to~\cite{Lee2001AutoMapping}.
Note that the dirty bit mask is only in the PIM L1 cache; processor caches
remain unchanged. Whenever a WAW is detected, the processor's copy of the cache
line is sent to the PIM core undergoing commit, which uses its dirty bit mask
to perform a merge.  As with any conventional multithreaded application, if the
programmer wishes to enforce a specific ordering of the two writes, they must
insert a synchronization primitive, such as a write fence (see
Section~\ref{sec:mech:synch}). }

\begin{figure}[h]
\centering
\vspace{-5pt}
\includegraphics[width=0.68\linewidth]{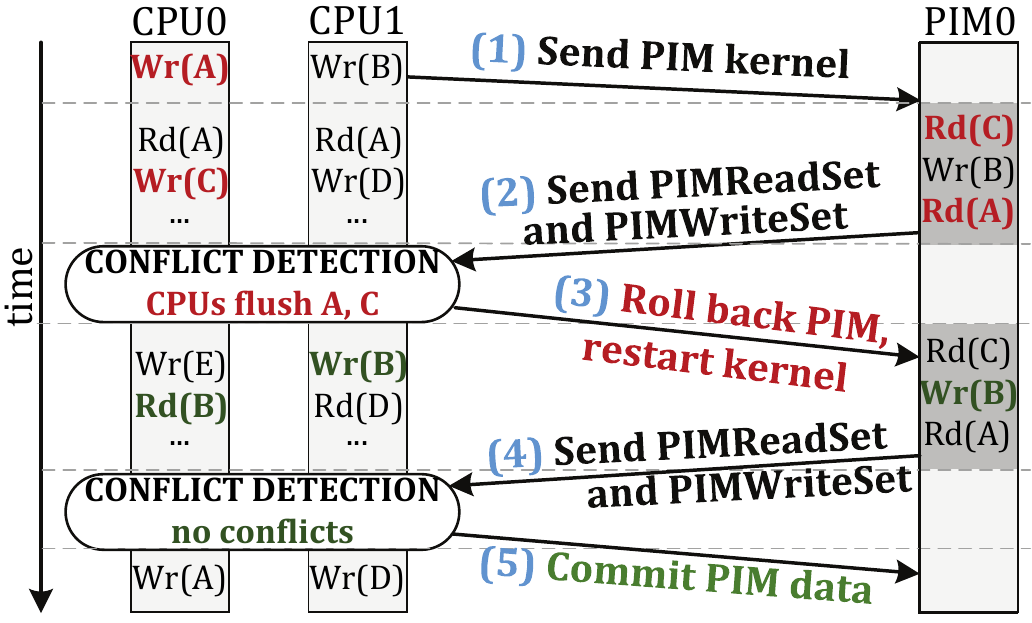}%
\vspace{-10pt}%
\caption{Example timeline of LazyPIM coherence.}
\label{fig:timeline}
\vspace{-8pt}
\end{figure}

\paratitle{Example.}
Figure~\ref{fig:timeline} shows an example timeline, where a PIM kernel is launched on PIM
core PIM0 while execution continues on processor cores CPU0 and CPU1.
As we mentioned previously, PIM kernel execution in LazyPIM behaves as if
\emph{the entire kernel's memory accesses} take place at the moment coherence is checked,
\emph{regardless of the actual time at which the kernel's accesses are
performed}.  Therefore, for \emph{every} cache line read by PIM0,
if CPU0 or CPU1 modify the line before the coherence
check occurs, PIM0 unknowingly uses stale data, leading to incorrect execution.
Figure~\ref{fig:timeline} shows
two examples of this: (1)~CPU0's write to line~C \emph{during} kernel execution; and
(2)~CPU0's write to line~A \emph{before} kernel execution, which was not 
written back to DRAM. 


Due to the use of coarse-grained atomicity in LazyPIM, even if the PIM kernel
writes to a cache line that is subsequently read by the 
processor before the PIM kernel finishes (such as the second PIM kernel write 
to line~B in Figure~\ref{fig:timeline}), this is \emph{not} a conflict.  The 
PIM kernel write is speculatively held in the PIM core cache, and does not 
commit until after the PIM kernel finishes.  Therefore, the processor read
\emph{effectively} takes place \emph{before} the PIM kernel write,
and thus does not read stale data.  Likewise, there is no conflict when CPU1
and PIM0 both write to line~B.

\subsection{Partial Kernel Commits} 
\label{sec:mech:partial}

One issue during LazyPIM conflict resolution is the overhead required to
rollback and restart the PIM kernel. If we wait until the end of kernel
execution to check coherence and attempt to commit PIM memory updates, the
probability of a conflict increases, as both the processor and the PIM cores
may have issued a large number of read and write operations. For applications
with high amounts of data sharing, this can lead to frequent rollbacks. While
we bound the maximum number of rollbacks that are possible for each kernel (see
Section~\ref{sec:arch:detect}), we still incur the penalty of re-executing the
\emph{entire} kernel, even if the conflict did not take place in the early
stages of kernel execution. Signature compression further exacerbates the
problem by introducing false positives during the conflict detection process
(Section~\ref{sec:arch:signature}), which increases the probability of a
rollback taking place.

\begin{figure}[h]
\centering
\vspace{-5pt}
\includegraphics[width=0.75\linewidth]{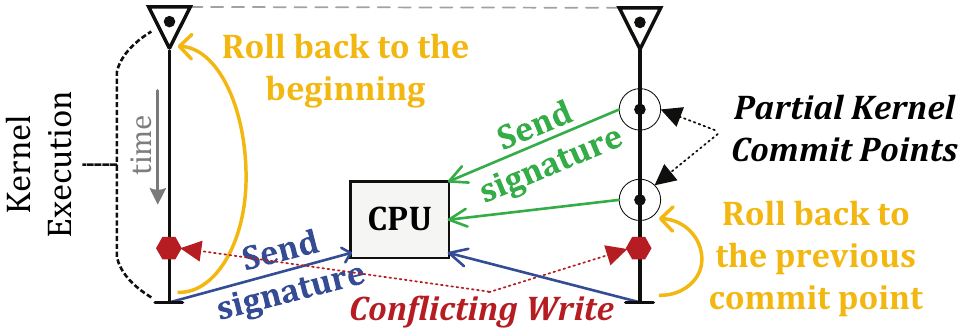}%
\vspace{-10pt}
\caption{PIM kernel execution without (left) and with (right) partial kernel commit support.}
\label{fig:partial_commit}
\vspace{-5pt}
\end{figure}

To reduce the probability of rollbacks occurring, we divide each PIM kernel
into smaller chunks of execution, (which we call \emph{partial kernels}), and
perform a commit after each partial kernel completes, as shown in
Figure~\ref{fig:partial_commit}.  Employing partial kernel commits in LazyPIM
offers three key benefits.  First, we lower the probability of conflicts, as we
now speculate for a shorter window of execution with fewer read and write
operations.  Second, we reduce the cost of performing a rollback, as the PIM
kernel needs to roll back only to the last commit point instead of to the
beginning of the kernel.  Third, we can reduce \emph{both} the size of the
signature and the false positive rate (see Section~\ref{sec:arch:signature}),
as the signature now needs to keep track of a much smaller set of addresses.
The use of partial kernel commits continues to follow the coherence model we
discussed earlier in the section, with all read and write operations for the
partial kernel treated as if they all occur at the moment that the partial
kernel commits.

One tradeoff of employing partial kernel commits is the fact that we must now
send coherence updates back to the processor more frequently.  If we send
coherence updates too often, we can undermine the bandwidth savings that
LazyPIM provides.  Ultimately, the effectiveness of partial kernel commit is
dependent on the length of each partial kernel.  We use two parameters to
control the length of the partial kernel.  First, a partial kernel is 
considered to be ready for commit if it performs a fixed number of memory
accesses.  This is motivated by our use of Bloom filters to implement
our compressed signatures, which can saturate if too many entries are inserted
into the filter (see Section~\ref{sec:arch:signature}).  Second, for partial kernels with very low memory access
rates, we also cap the number of instructions that can be executed by any one
partial kernel.  That way, if a rollback needs to occur, we provide an upper
bound on the performance impact that rollback can have.  We discuss these
mechanisms for partial kernel sizing in Section~\ref{sec:arch:partial}.

\subsection{Sharing Between PIM Kernels} 
\label{sec:mech:multicore}

\changes{LazyPIM allows data sharing between PIM cores to simplify the user
programming model, to allow transparent communication between PIM cores, and to
improve overall PIM performance.} When multiple PIM kernels run together, the
kernels execute on multiple PIM cores whose L1 caches are kept coherent with
each other through the use of a local PIM directory, as we described in
Section~\ref{sec:bkgd:baseline}. \changes{While we observe no sharing of
speculative data between PIM cores in any of our workloads, LazyPIM includes
hardware mechanisms to allow the PIM cores to interact with each other safely.
Our mechanism ensures} that (1)~the propagation of speculative data between PIM
kernels is tracked; and (2)~if one PIM kernel rolls back, any other uncommitted
PIM kernel that read data produced by the rolled back kernel also rolls back.
\changes{ Note that LazyPIM ensures that livelock does not occur,
guaranteeing forward progress (see Section~\ref{sec:arch:detect}).}

LazyPIM ensures that the sharing of (potentially stale) data between multiple PIM kernels does not lead
 to incorrect execution. Each PIM core maintains a per-core bit vector (which we call \emph{speculative read bits}).
Whenever a PIM core reads speculative data written by another PIM core, it sets the corresponding speculative
 read bit for the ID of the PIM core that it read from. When the PIM core is ready to commit its kernel,
the core checks to see whether any speculative read bits are set.
If set, the core waits to perform its conflict detection until all of the PIM cores from which it read
 speculatively also complete their kernels.  During conflict detection, if \emph{all} of the signatures indicate no conflict,
 then the commit procedure continues for all of the PIM cores that shared
data. If a conflict is detected in \emph{any} of the cores' signatures, then \emph{all} of the cores roll back. 
In essence, whenever PIM cores share speculative data with each other, their memory operations are grouped
 together so that they all take place atomically, behaving
as a single PIM kernel.  We made this design choice to keep the design simple.

If no speculative data exchange takes place between PIM kernels, then each PIM kernel can commit
 independently without having to worry about any coordination. If two independent PIM kernels try to commit at the same time, the commit process serializes
the commit requests.

\subsection{Support for Synchronization Primitives} 
\label{sec:mech:synch}

\changes{LazyPIM supports synchronization primitives and atomic operations. To
guarantee atomicity and avoid ordering violations, LazyPIM forces a partial
commit when a core reaches a synchronization primitive. The partial commit
checks coherence permissions and ensures that all updates from the PIM core
are immediately visible globally. By using partial commits, LazyPIM
effectively provides coherence at a cache line granularity
for synchronization primitives.

Let us look at the Acquire/Release primitives as an example. When a processor thread acquires a lock, it reads and updates a
shared variable associated with the lock. If a PIM kernel attempts to acquire
this lock, a partial commit forces the PIM kernel to read the updated value
written by the processor thread, thus
guaranteeing mutual exclusion. Once the processor thread has freed the lock,
the PIM kernel can acquire the lock by writing to the shared variable,
immediately performing a partial commit to make the lock acquisition globally
visible. This ensures that the processor immediately sees the updated value of
the shared variable. Note that during the partial commit, the processor cannot
inadvertently update the shared value, as the partial commit mechanism locks
all memory in the PIM data region.}




\section{LazyPIM Architecture}
\label{sec:arch}

In this section, we discuss the hardware support
 LazyPIM provides for speculative execution and coherence conflict 
detection.
Using simple programmer annotations (Section~\ref{sec:arch:program}), LazyPIM
tracks the data that \emph{might} be accessed by the PIM kernels 
during speculative execution in hardware (Section~\ref{sec:arch:spec}).
LazyPIM employs compressed signatures (Section~\ref{sec:arch:signature}) to 
record the accessed data until it detects that the partial kernel has finished
executing (Section~\ref{sec:arch:partial}).  Once the partial kernel finishes,
LazyPIM performs \emph{signature-based conflict detection} 
(Section~\ref{sec:arch:detect}).  In order to reduce the probability of conflicts,
LazyPIM tracks and proactively writes back dirty PIM data sitting in the processor
cache by employing a variant of the Dirty-Block Index~\cite{seshadri2014dirty}
(Section~\ref{sec:arch:pimdbi}).  We discuss the hardware overhead of
LazyPIM in Section~\ref{sec:arch:overhead}.

\subsection{Program Interface}
\label{sec:arch:program}

We provide a simple 
interface
for programmers
to port applications to 
LazyPIM. 
We show the implementation of a simple LazyPIM kernel within a program in
Listing~\ref{code:program_lazypim}. 
The programmer identifies the portion(s) of
the code to execute on PIM cores, 
using two macros (\texttt{PIM\_begin} and \texttt{PIM\_end}).
The compiler converts the macros into instructions
that we add to the ISA, which \emph{trigger} and \emph{end} PIM kernel execution. 
LazyPIM also needs to know which parts of the 
allocated 
data \emph{might} be accessed by the PIM cores,
which we call the \emph{PIM data region}.
We assume either the
programmer or the compiler can annotate all of the PIM
data using compiler directives or a PIM memory allocation API. This information is stored in the
page table using per-page flag bits, indicating that a cache line might be
accessed by a PIM kernel. 

    \begin{lstlisting}[frame=single, language=C, captionpos=b, caption=Example PIM program implementation.,
label=code:program_lazypim, basicstyle=\small,
numbers=left, xleftmargin=1.75\parindent, xrightmargin=-1.75\parindent, framexrightmargin=-2.1\parindent, basewidth=0.58em,
float,floatplacement=H]
PageRank(@PIM Graph GA) {
  @PIM double* p_curr, p_next;
  @PIM bool* frontier;
  vertexSubset Frontier(n, n, frontier);
  ...
  while(iter++ < maxIters) {
    PIM_BEGIN
    vertexSubset output = edgeMap(GA, Frontier, 
      PR_F<vertex>(p_curr, p_next, GA.V), 0);
    PIM_END
    vertexMap(Frontier, PR_Vertex_F); 
  ...
\end{lstlisting} 



\subsection{Speculative Execution}
\label{sec:arch:spec}

\begin{figure}[h]
\centering
\includegraphics[width=0.8\linewidth]{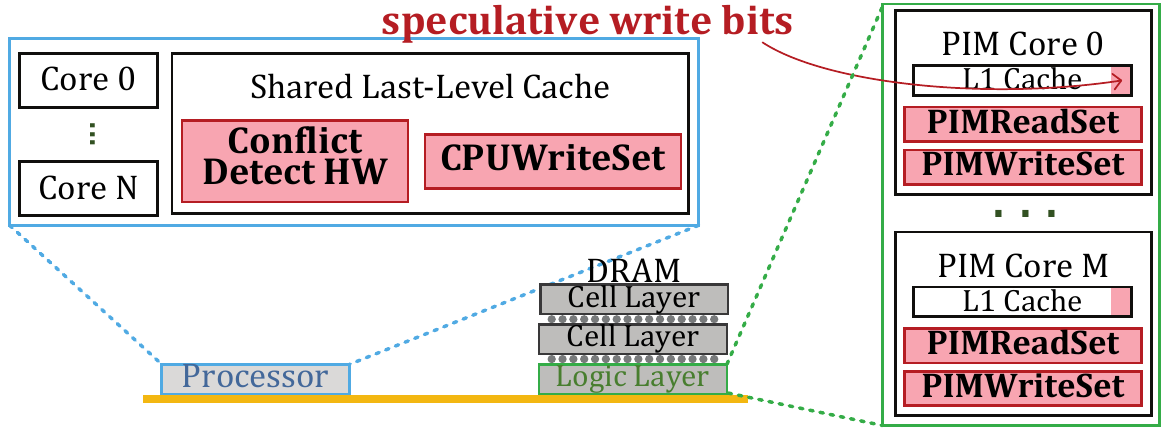}%
\caption{High-level additions (in bold) to PIM architecture to support LazyPIM.}
\label{fig:additions}
\vspace{-10pt}
\end{figure}

When an application reaches a \emph{PIM kernel trigger} instruction, the processor
dispatches the kernel's starting PC and live-in registers to a free PIM core.
The PIM core makes a \emph{checkpoint} at the starting PC, in case a rollback is required later, 
and starts executing the kernel.  The kernel 
\emph{speculatively} assumes that it has coherence permissions for \emph{every} 
cache line it accesses, without \emph{actually} 
checking the processor directory.  We add 
a one-bit flag to each cache line in the PIM core cache, as shown in Figure~\ref{fig:additions},
to mark all data updates as speculative. 
If a speculative cache line is selected for eviction, the PIM core stops PIM
kernel execution and attempts a partial kernel commit (see 
Section~\ref{sec:arch:partial}).

LazyPIM tracks three sets of cache line addresses during PIM kernel
execution.  These sets are recorded into three \emph{signatures}
(explained in detail in Section~\ref{sec:arch:signature}), as shown in
Figure~\ref{fig:additions}: 
(1)~the \cpuwriteset (all \emph{CPU writes} to the PIM data region), 
(2)~the \pimreadset (all \emph{PIM core reads}), and 
(3)~the \pimwriteset (all \emph{PIM core writes}).
When a partial kernel starts, the
processor scans the tag store and records all dirty cache lines in
the PIM data region (identified using the page table flag bits from Section~\ref{sec:arch:program}) in the
\cpuwriteset. During PIM kernel execution, the processor updates the
\cpuwriteset whenever it writes to a cache line in the PIM data
region. In the PIM core, the \pimreadset and \pimwriteset are updated for
\emph{every} read and write performed by the partial PIM kernel.  When
the partial PIM kernel finishes execution, the three signatures are
used to resolve speculation (see Section~\ref{sec:arch:detect}).

\subsection{Signatures}
\label{sec:arch:signature}

LazyPIM uses fixed-length parallel Bloom
filters~\cite{Bloom:1970:STH:362686.362692} to implement the compressed
signatures.  The signatures store data by partitioning an $N$-bit signature
into $M$ segments (such that each segment is $N$/$M$ bits wide).  To add an
address into the signature, each of the $M$ segments employs a unique hash
function ($H_{3}$~\cite{Sanchez:2007:IST:1331699.1331713})  that maps the address to a single bit within the segment, which we
call the \emph{hashed value}.  To check whether an address is present within
the signature, we generate the $M$ hashed values of that address and then check
to see if every corresponding bit is set in each segment.  Bits that are set in
the signature remain set until the signature is reset, preventing any false
negatives from occurring.  However, there may be some collision for the bits
set by various hashed values, which can lead to a limited number of false
positives.

Parallel Bloom filters allow us to easily determine if any conflicts occurred.
We can quickly generate the intersection of two signatures (i.e., a list of
hash values that exist in \emph{both} signatures) by taking the bitwise AND of
the two signatures.  If we find that any of the $M$ segments in the
intersection are empty, no conflicts exist between the two signatures.
Otherwise, the intersection contains the hashed values of all addresses that
might exist in both signatures (including potential false positives).

\paratitle{Signature Size.}
Bloom filter based signatures allow for a large number of addresses to be stored
within a single fixed-length register.
The size of the signature is
directly correlated with both the number of addresses we want to store and the
desired false positive rate.  A larger signature leads to fewer false positives
and therefore fewer rollbacks, which can improve performance.  However, larger
signatures require greater storage overhead and require more bandwidth when we
send the \pimreadset and \pimwriteset to the processor for conflict detection
(see Section~\ref{sec:arch:detect}). Therefore, our goal is to select a
signature width that provides a balance between rollback minimization and
bandwidth utilization.

For the \pimreadset and \pimwriteset, we use one 2~Kbit register for each
signature with $M$ set to 4.  We discuss how the signature size is used to
control the length of partial kernels in Section~\ref{sec:arch:partial}. We can
employ \emph{multiple} 2~Kbit registers for the \cpuwriteset in order to increase
the number of addresses that the signature can hold since the \cpuwriteset does
not need to be transmitted off-chip (see Section~\ref{sec:arch:detect}).  We
find that expanding the \cpuwriteset capacity is helpful, as the
\cpuwriteset records both the processor writes that occur during PIM kernel
execution and the dirty cache lines that reside in the cache \emph{at the time
partial kernel execution begins}.  Note that we must keep each register at the
same 2~Kbit width (with $M=4$) as the \pimreadset and \pimwriteset so that we
can perform an intersection test between the two. For every address stored in the
\cpuwriteset, we use round robin selection to choose which of the registers the
hashed value of the address is written to.  During conflict detection, we
intersect the \pimreadset and \pimwriteset with \emph{each} of the registers in
the \cpuwriteset.

\subsection{Support for Partial Kernels}
\label{sec:arch:partial}

To maximize the effectiveness of partial kernel commits (see
Section~\ref{sec:mech:partial}), we must ensure that the \pimreadset and
\pimwriteset do not get saturated with too many addresses (which can lead to a
high rate of false positives).  LazyPIM avoids this by dynamically determining
the size of each partial kernel.  We determine the maximum number of addresses
that we allow inside the \pimreadset and \pimwriteset, which we compute based on
our desired false positive rate.  One property of Bloom filters is that they can 
guarantee a maximum false positive rate for a given filter width and a set 
number of address insertions.  To support this, we maintain one counter each
for the \pimreadset and \pimwriteset of each PIM core, which is incremented every
time we insert a new address.  When we reach the maximum number of addresses for
either signature, we stop the partial kernel and initiate the conflict detection
process.  In our implementation, we target a 30\% false positive rate, which
allows us to store up to 250 addresses within a 2~Kbit signature.

One potential drawback, as mentioned in Section~\ref{sec:mech:partial}, is that
for PIM kernels with a low memory access rate, each partial kernel can consist
of a large number of instructions, increasing the cost of rollback in the event
of a conflict.  To avoid the costly rollback overhead, we employ a third 
counter for each PIM core, which counts the number of instructions executed for
the current partial kernel.  If this counter exceeds a fixed threshold (set in
our implementation empirically to 1~million instructions), we also
stop the partial kernel, even if neither the \pimreadset nor the \pimwriteset are
full.

\subsection{Handling Conflicts}
\label{sec:arch:detect}

As discussed in Section~\ref{sec:mech:conflicts},
we need to detect conflicts that occur during PIM kernel execution,
such as the PIM0 reads to lines~A and~C in Figure~\ref{fig:timeline}.  
In LazyPIM, when the partial kernel finishes execution, both the \pimreadset 
and \pimwriteset are sent back to the processor.  We then compute the 
intersection of the \pimreadset and \cpuwriteset to determine if any 
addresses exist in both signatures.

If no matches are detected between the \pimreadset and the \cpuwriteset (i.e.,
no conflicts have occurred), the partial kernel \emph{commit} starts.
\changes{First, the intersection of the \pimwriteset and \cpuwriteset is
computed. If any matches are found, the conflicting cache lines are sent to the
PIM core to be merged with the PIM core's copy of the cache line (as discussed
in Section~\ref{sec:mech:conflicts}).} \changes{Second, any clean cache lines}
in the processor cache that match an address in the \pimwriteset are
invalidated. A message is then sent to the PIM core, allowing it to write its
speculative cache lines back to DRAM. During the commit, all coherence
directory entries for the PIM data region are locked to ensure atomicity.
Finally, all signatures are erased, and the next partial kernel begins. 

If an overlap is found between the \pimreadset and the \cpuwriteset,
a conflict \emph{may} have occurred.
The processor flushes only the dirty cache lines that match 
addresses in the \pimreadset back to DRAM.
During the flush, all
PIM data directory entries are locked to ensure atomicity.  Once 
the flush completes, a message is sent to the PIM core, telling it to
invalidate all speculative cache lines, and to \emph{roll back} to the
beginning of the partial kernel using the checkpoint.
All signatures are erased, and the PIM core restarts partial kernel execution.
After
re-execution finishes, conflict detection is performed again.

\changes{Note that during the commit process, processor cores do not stall
unless they access the same data accessed by a PIM core.} LazyPIM guarantees
forward progress by acquiring a lock for each line in the \pimreadset after a
number of rollbacks \changes{(we empirically set this number to 3~rollbacks).
This simple mechanism ensures there is no livelock even if the sharing of
speculative data amongst PIM cores might create a cyclic dependency. Note that
rollbacks are caused by CPU accesses to conflicting addresses, and not by the
sharing of speculative data between PIM cores. As a result, once we lock
conflicting addresses following 3 rollbacks, the PIM cores will not rollback
again as there will be no conflicts, guaranteeing forward progress.}

\subsection{Dirty-Block Index for PIM Data}
\label{sec:arch:pimdbi}

One issue that exacerbates the number of conflicts in LazyPIM is the need to
record dirty PIM data within the processor cache as part of the \cpuwriteset,
as any changes not written back to memory will not be observed by the PIM
cores, even if the changes took place before the PIM kernel started (we refer
to these as \emph{dirty conflicts}).  We analyze the impact of dirty conflicts,
and make two key observations.  First, across our workloads, majority of conflicts 
that occur are a result of dirty conflicts.
Second, we observe that an average of 95.4\% of the addresses inserted into the \cpuwriteset
are due to dirty conflicts.  If we were able to reduce the dirty conflict 
count, we could significantly reduce the overhead of conflict detection, and
could reduce the probability of rollbacks.

In LazyPIM, we optimize the dirty conflict count by introducing a Dirty-Block
Index~\cite{seshadri2014dirty}.  The Dirty-Block Index (DBI) reorganizes the
tag store of a cache to track the dirty cache lines of memory pages, which it
then uses to opportunistically write back dirty data during periods of low
cache and bandwidth utilization.  LazyPIM employs what we call the PIM-DBI in
the processor, which is in essence a DBI dedicated solely for the PIM data 
region.  To simplify the implementation of DBI in LazyPIM, we maintain a cycle
counter that triggers PIM-DBI at fixed intervals.  Whenever PIM-DBI is 
triggered, it writes back dirty cache lines within the PIM data
region to DRAM, which reduces the number of dirty conflicts that
occur at the time a PIM kernel starts execution.  Note that LazyPIM does not require the
PIM-DBI.

\subsection{Hardware Overhead}
\label{sec:arch:overhead}

Each signature register in LazyPIM is 2~Kbits wide, with the \pimreadset and
\pimwriteset consisting of a single register, while the \cpuwriteset includes
16~registers.  Overall, each PIM core uses 512B for signature storage, while
the processor uses 8KB in total.  
Aside from the signatures, LazyPIM's overhead consists mainly of 
(1)~1~bit per page (0.003\% of DRAM capacity) and 1~bit per TLB entry for the page table
flag bits (Section~\ref{sec:arch:program}); 
(2)~a 0.2\% increase in PIM core L1 size to mark speculative data (Section~\ref{sec:arch:spec}); 
\changes{(3)~a 1.6\% increase in PIM core L1 size for the dirty bit mask (Section~\ref{sec:mech:conflicts});} and 
(4)~two 8-bit counters and one 20-bit counter in each PIM core to track the 
number of addresses and instruction count for partial kernels (Section~\ref{sec:arch:partial}).

To track speculative data sharing between PIM cores (Section~\ref{sec:mech:multicore}),
we also require $P-1$~bits in each PIM core for a system with $P$ PIM cores.
For each PIM core, this comes to a total of 596~bytes of overhead in a system with 16 PIM cores.
The PIM-DBI tag store (Section~\ref{sec:arch:pimdbi}) is sized to track 1024 cache blocks, split into rows of 64 blocks each.  A single row stores a 64-bit dirty
bit array and a 48-bit tag, resulting in a total storage overhead of 224B.  

\section{Methodology}
\label{sec:methodology}

In this section, we discuss our methodology for evaluating LazyPIM.
We study two types of data-intensive applications as case studies: graph 
workloads and in-memory databases (Section~\ref{sec:methodology:apps}).
We identified ideal candidates for PIM kernels within each application using a
profile-driven procedure (Section~\ref{sec:methodology:kernels}).
We used the full-system version of the gem5 simulator~\cite{GEM5} to perform
quantitative evaluations (Section~\ref{sec:methodology:simulator}).

\subsection{Applications}
\label{sec:methodology:apps}

Ligra~\cite{ligra} is a lightweight multithreaded graph processing system for
shared memory multicore machines. 
We use three Ligra graph applications (PageRank, Radii, and Connected Components), with input
graphs constructed from real-world network datasets~\cite{snap}: Enron email communication network (73384 nodes, 367662 edges), arXiV General Relativity (10484 nodes, 28984 edges), and peer-to-peer Gnutella25 (45374 nodes, 109410 edges).


We also use an in-house prototype of an in-memory database (IMDB) that is capable
of running both transactional queries (similar to TPC-C) and analytical queries (similar
to TPC-H) on the same 
database tables, and represents modern IMDBs~\cite{SAP,stonebraker2013voltdb,MemSQL}
that support HTAP workloads.
Our transactional workload consists of 64K
 transactions, with each transaction performing reads or writes on a few randomly-chosen
database tuples. Our three analytical workloads consist of \changes{128, 192, or 256} analytical queries 
(for \emph{HTAP-128}, \changes{\emph{HTAP-192},} and \emph{HTAP-256}, respectively) that
 use the select and join operations. The IMDB uses a
state-of-the-art, highly-optimized hash join kernel code~\cite{Teubner:2013:MHJ:2510649.2511320}
 tuned for IMDBs to service the join queries.
We simulate an IMDB system that has
64 tables, where each table consists of 64K tuples, and each tuple has 32 fields. Tuples in tables are populated by a randomly generated
uniformly-distributed integer.






\subsection{Identifying PIM Kernels}
\label{sec:methodology:kernels}

We select PIM kernels from our applications with the help of OProfile~\cite{OProfile}.
Our goals during partitioning were to lower execution time and minimize data movement,
 which all PIM mechanisms aim to
achieve~\cite{Ahn:2015:PIL:2749469.2750385, 7056040, Mingyu:PACT}.
 As prior work~\cite{Ahn:2015:PIL:2749469.2750385} has shown, 
parts of the application that are either (1)~compute-intensive or
 (2)~cache-friendly
should remain on the processor, as they can benefit from larger,
 more sophisticated cores with larger caches (as opposed
to the small PIM cores). Thus, only memory-intensive portions of
 the code with poor cache locality are dispatched to the PIM
cores for execution. To partition an application, we profiled its
 execution time, along with the miss rate for each function, using a
training set input. Based
on the profiling information, we selected candidate PIM
 portions conservatively, choosing portions where (1)~the application
spent a majority of its cycles (to amortize the overhead of launching
 PIM execution), and (2)~a majority of the total last-level
cache misses occurred (indicating high memory intensity and poor
 cache locality). From this set of candidate kernels, we selected
only the kernels for each mechanism that had minimal coherence
 overhead under a given mechanism (i.e., they minimized the
communication that occurred between the processor threads and
 the PIM kernels).


We modify each application to ship the selected kernels to
the PIM cores.  We manually annotated the PIM data set such that
gem5 can distinguish between data that belongs to the PIM data region from data that does not. To ensure correct annotation, 
we tracked all of the data used by any PIM kernel, and replaced the
 \texttt{malloc} for all of this data with a specialized memory allocation function that we wrote, called \texttt{pim\_alloc}.
Our \texttt{pim\_alloc} function automatically registers the allocated data as belonging to the PIM data region within gem5.

\subsection{Simulation Environment}
\label{sec:methodology:meth}
\label{sec:methodology:simulator}

We use the gem5~\cite{GEM5} architectural simulator in full-system mode,
using the x86 ISA, to implement our proposed coherence mechanism.
DRAMSim2~\cite{DRAMSim2} was used within gem5 to provide detailed
DRAM timing behavior.  We modified the DRAM model to emulate the high in-memory
bandwidth available within HMC~\cite{hmcspec2} for the PIM cores.


\begin{table}[h]
    \small
    \vspace{-3pt}
    \renewcommand{\arraystretch}{0.5}
    \setlength{\aboverulesep}{1pt}
    \setlength{\belowrulesep}{2pt}

    \centering
    \label{tbl:config}%
    \setlength{\tabcolsep}{.5em}
    \begin{tabular}{ll}
        \toprule
        \emph{Processor} &  
        4--16 cores, 8-wide issue, 2 GHz frequency\\
        & \emph{L1 I/D Caches}:    
        64kB private, 4-way associative, 64B blocks \\
        & \emph {L2 Cache}
        2MB shared, 8-way associative, 64B blocks\\
        & \emph{Coherence}: MESI \\
        \midrule
        \emph{PIM} & 
        4--16 cores, 1-wide issue, 2 GHz frequency \\
        & \emph{L1 I/D Caches}: 64kB private, 4-way associative, 64B blocks\\
        & \emph{Coherence}: MESI \\
        \midrule
        \emph{HMC~\cite{hmcspec2}} & 
              one 4GB cube, 16 vaults per cube, 16 banks per vault \\
        \midrule
        \emph{Memory} &
              DDR3-1600, 4GB, FR-FCFS scheduler \\
        \bottomrule
    \end{tabular}%

    \caption{Evaluated system configuration.}
\end{table}

We model the sum of the energy consumption that takes place within the DRAM, off-chip interconnects, and all caches.
We also include DBI energy consumption and the energy used by other components of LazyPIM in our energy model. 
We model DRAM energy as the energy consumed per bit, leveraging estimates from prior work~\cite{6242474}.
We estimate the energy consumption of 
all L1 and L2 caches using CACTI-P 6.5~\cite{CACTI}, assuming a 22nm process.  
We model the off-chip interconnect using the method used by prior work~\cite{Mingyu:PACT}, which estimates the HMC SerDes energy consumption as 3pJ/bit for data packets.

\section{Evaluation}
\label{sec:eval}

We examine how LazyPIM compares with prior coherence
mechanisms for PIM.
Unless otherwise stated, we assume that LazyPIM employs partial kernel commits
as described in Section~\ref{sec:arch:partial}, and that it uses PIM-DBI, where
all dirty \changes{PIM} data in the processor caches is written back every 800K processor cycles.
We show results normalized to a processor-only baseline (\emph{CPU-only}),
and compare to PIM execution using fine-grained coherence (\emph{FG}), coarse-grained locks 
(\emph{CG}), or non-cacheable PIM data (\emph{NC}),
as described in Section~\ref{sec:motivation:naive}.


\subsection{Performance}
\label{sec:eval:perf}

\begin{figure}[h]
    \centering
    \vspace{-5pt}
    \includegraphics[width=\linewidth]{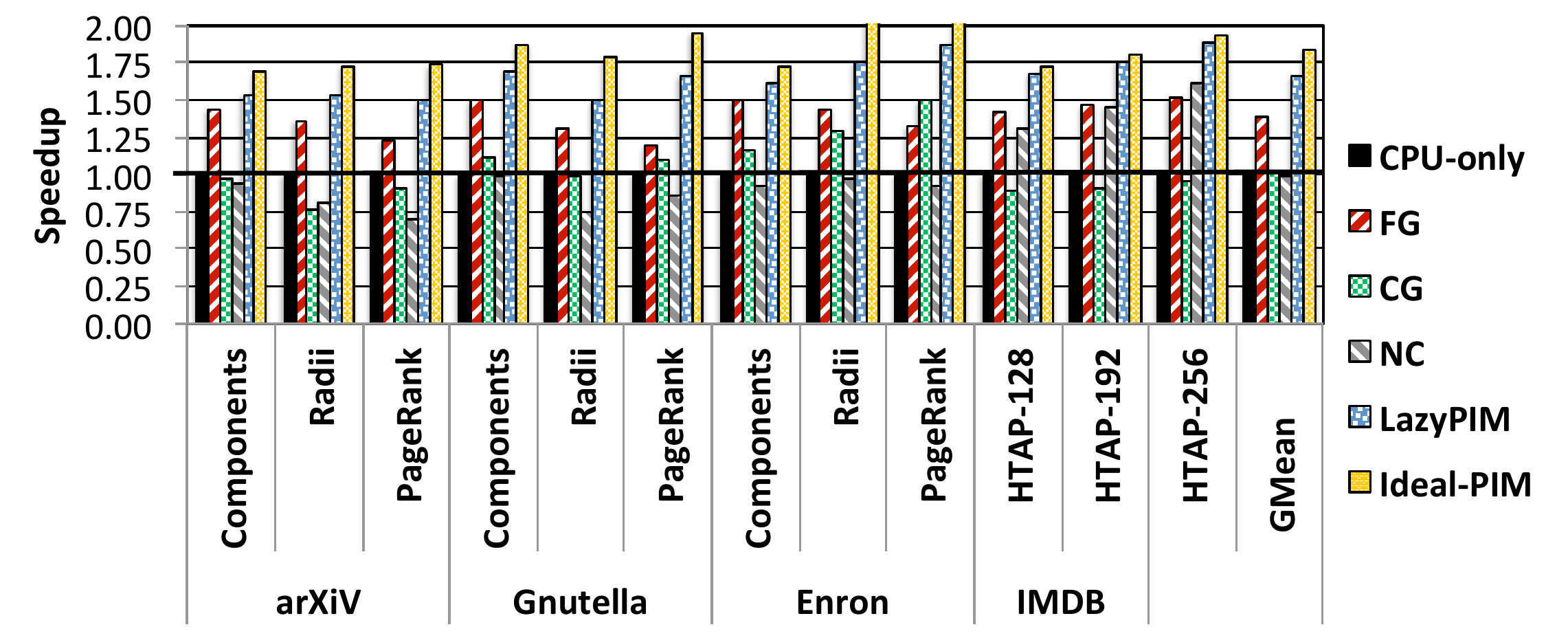}%
    \vspace{-8pt}%
    \caption{Speedup for all applications with 16~threads, normalized to CPU-only execution.}
    \label{fig:performance2}
    \vspace{-5pt}
\end{figure}

Figure~\ref{fig:performance2} shows the performance improvement for a 16-core 
architecture (with 16 processor cores and 16~PIM cores) across all of our 
applications and input sets.
Without any coherence overhead, the ideal PIM mechanism (Ideal-PIM, as defined in Section~\ref{sec:motivation:naive}) 
significantly outperforms CPU-only across all applications, showing the potential of PIM execution on these workloads.
The poor handling of coherence by CG and NC, and by FG in a number of cases, leads to drastic performance losses compared to Ideal-PIM,
indicating that an efficient coherence mechanism is essential for PIM performance.
For example, in many cases, NC and CG actually perform \emph{worse} than CPU-only.
In contrast, LazyPIM consistently retains most of Ideal-PIM's benefits for all applications, coming within
\changes{9.8\%} of the Ideal-PIM performance, on average. 
 LazyPIM outperforms all other approaches, improving average performance by \changes{19.6\% over FG, 65.9\%
 over CG, 71.4\% over NC, and 66.0\% over CPU-only.} 

\begin{figure}[t]
    \centering
        \centering
        \includegraphics[width=\linewidth]{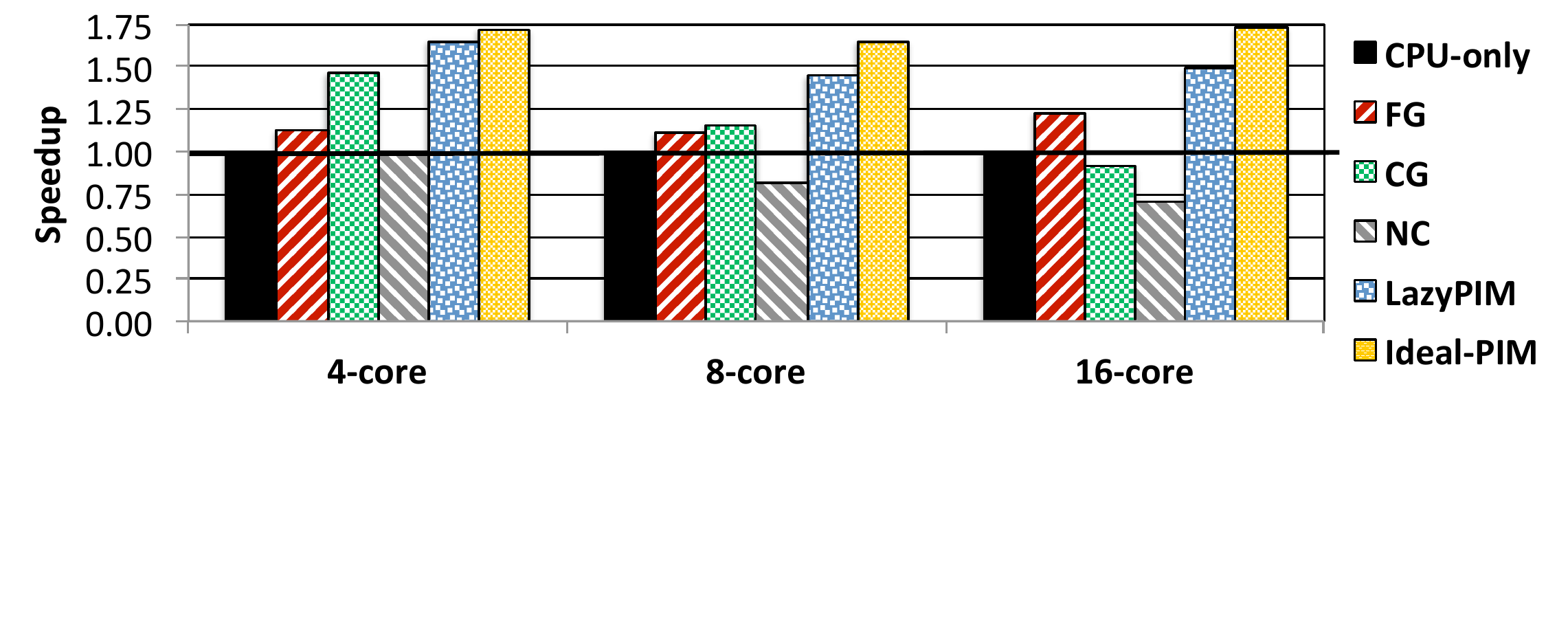}%
\vspace{-45pt}%
    \caption{Speedup vs.\ thread count for PageRank with arXiV graph, normalized to CPU-only execution.}
    \label{fig:performance1}
   \vspace{-2pt}
\end{figure}

We select PageRank using the ArXiv input set for a case study on how
performance scales as we increase the core count, shown in 
Figure~\ref{fig:performance1}.
 Ideal-PIM significantly outperforms CPU-only at all core counts. 
With NC, the processor threads
 incur a large penalty for going to DRAM frequently to access the PIM data region.
 CG suffers greatly due to (1)~flushing dirty cache lines, and (2)~blocking all processor
 threads that access PIM data during execution. In fact, processor threads 
 are blocked for up to 73.1\% of the total execution time with CG. With more core, the effects of 
 blocking worsen CG's performance. FG also loses a significant portion of Ideal-PIM's 
 improvements, as it sends a large amount of off-chip messages.
Note that FG, despite its high off-chip traffic, scales better with core count 
than CG and NC, as it neither blocks processor cores nor slows down CPU execution.
LazyPIM improves performance at all core counts.


We conclude that LazyPIM successfully harnesses the benefits of
speculative coherence updates to provide performance improvements in a wide
range of PIM execution scenarios, when prior coherence mechanisms cannot
improve performance and often degrade it.

\subsection{Off-Chip Traffic}
\label{sec:eval:offchip}

\begin{figure}[h]
    \vspace{-4pt}
    \centering
    \includegraphics[width=\linewidth]{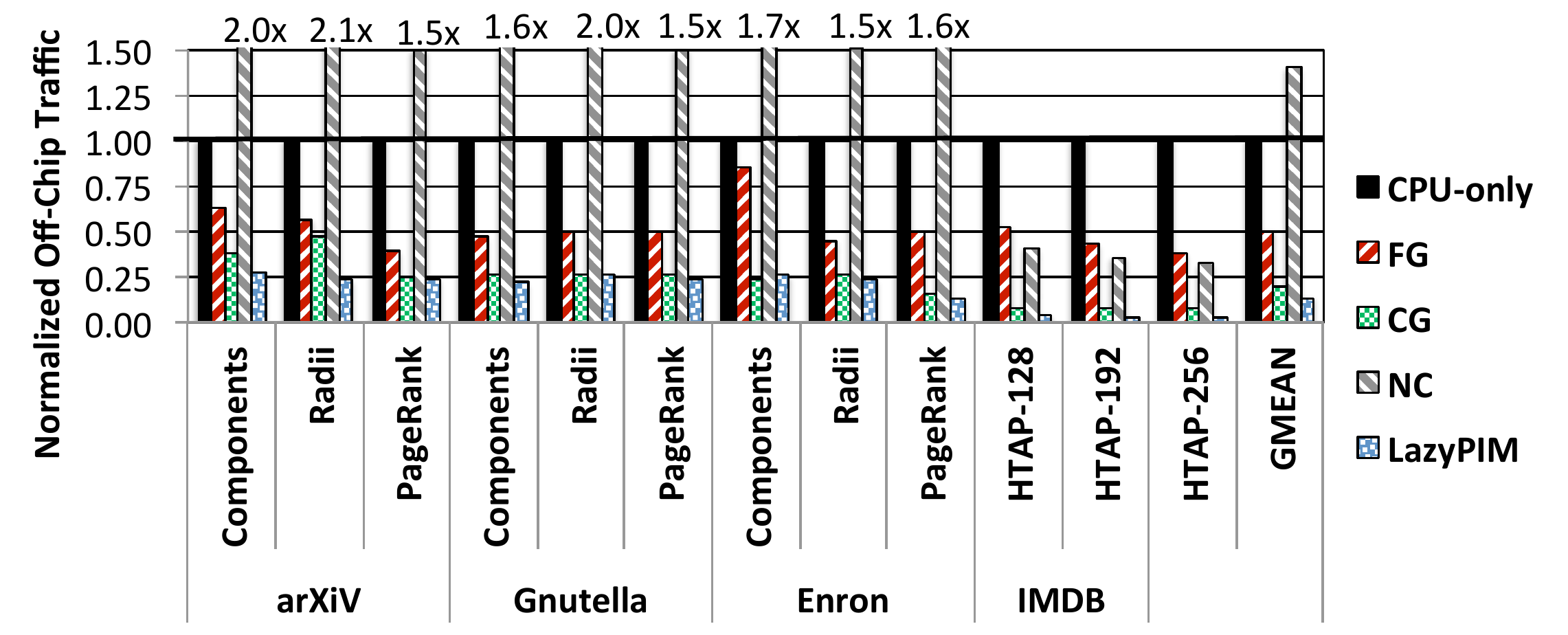}%
    \vspace{-8pt}%
    \caption{16-thread off-chip traffic, normalized to CPU-only execution. \emph{Note that lower is better.}}
    \label{fig:traffic2}
\vspace{-5pt}
\end{figure}

Figure~\ref{fig:traffic2} shows the normalized off-chip traffic of the PIM 
coherence mechanisms for our 16-core system.  LazyPIM significantly 
reduces \emph{overall} off-chip traffic, with an average reduction of \changes{30.9\%} 
over CG, the best prior approach in terms of traffic.
CG has greater traffic than LazyPIM mainly because CG
has to flush dirty cache lines before each PIM kernel invocation,
while LazyPIM takes advantage of speculation to perform \emph{only} the
\emph{necessary} flushes \emph{after} the PIM kernel finishes execution.
As a result, LazyPIM reduces the flush count (e.g., by 92.2\% for
Radii using arXiV), and thus lowers overall 
off-chip traffic (by \changes{50.8\%} for our example).
NC's traffic suffers from the fact that \emph{all} processor accesses to the
PIM data region must go to DRAM.
With respect to CPU-only, LazyPIM provides an average traffic reduction of 
\changes{86.3\%}.

\begin{figure}[h]
    \vspace{-7pt}
    \centering
    \includegraphics[width=\linewidth]{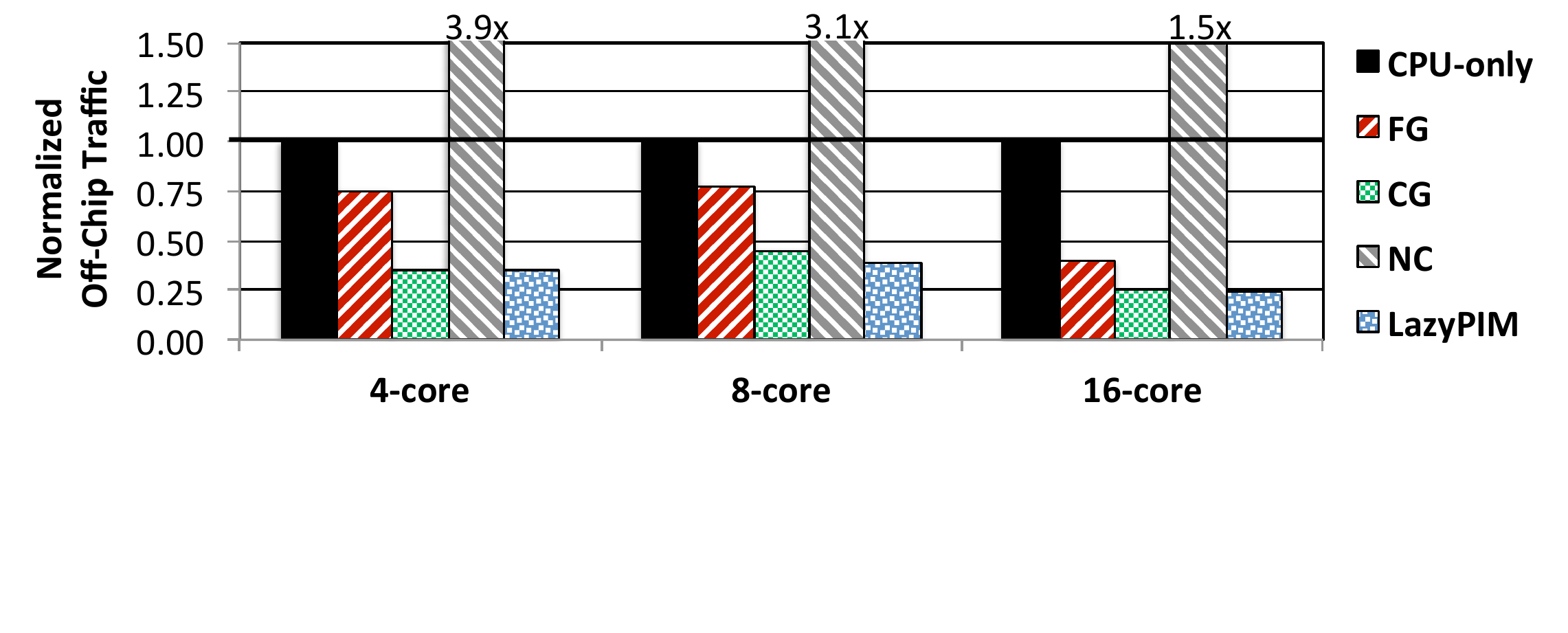}%
    \vspace{-45pt}%
    \caption{PageRank off-chip traffic sensitivity to thread count with arXiV graph, normalized to CPU-only execution. \emph{Note that lower is better.}}
    \label{fig:traffic1}
   \vspace{-7pt}
\end{figure}

Figure~\ref{fig:traffic1} shows the normalized off-chip traffic as the 
number of cores increases, for PageRank using the arXiV graph.  
LazyPIM's traffic scales better with core count than prior PIM 
coherence mechanisms. Due to false sharing, the number of flushes for CG scales \emph{superlinearly} with 
thread count (not shown), increasing 6.2x from 4 to 16~threads.
NC also scales poorly with core count, as more processor threads generate a greater number of accesses.  In
contrast, LazyPIM allows processor cores to cache PIM data, by enabling
coherence efficiently, which lowers off-chip traffic on average by \changes{88.3\%} with respect
to NC.




We conclude that the efficient approach to coherence employed by LazyPIM allows
it to successfully provide fine-grained coherence behavior while still 
delivering reductions in off-chip traffic.

\subsection{Energy}
\label{sec:eval:energy}

\begin{figure}[h]
    \centering
    \vspace{-5pt}
    \includegraphics[width=\linewidth]{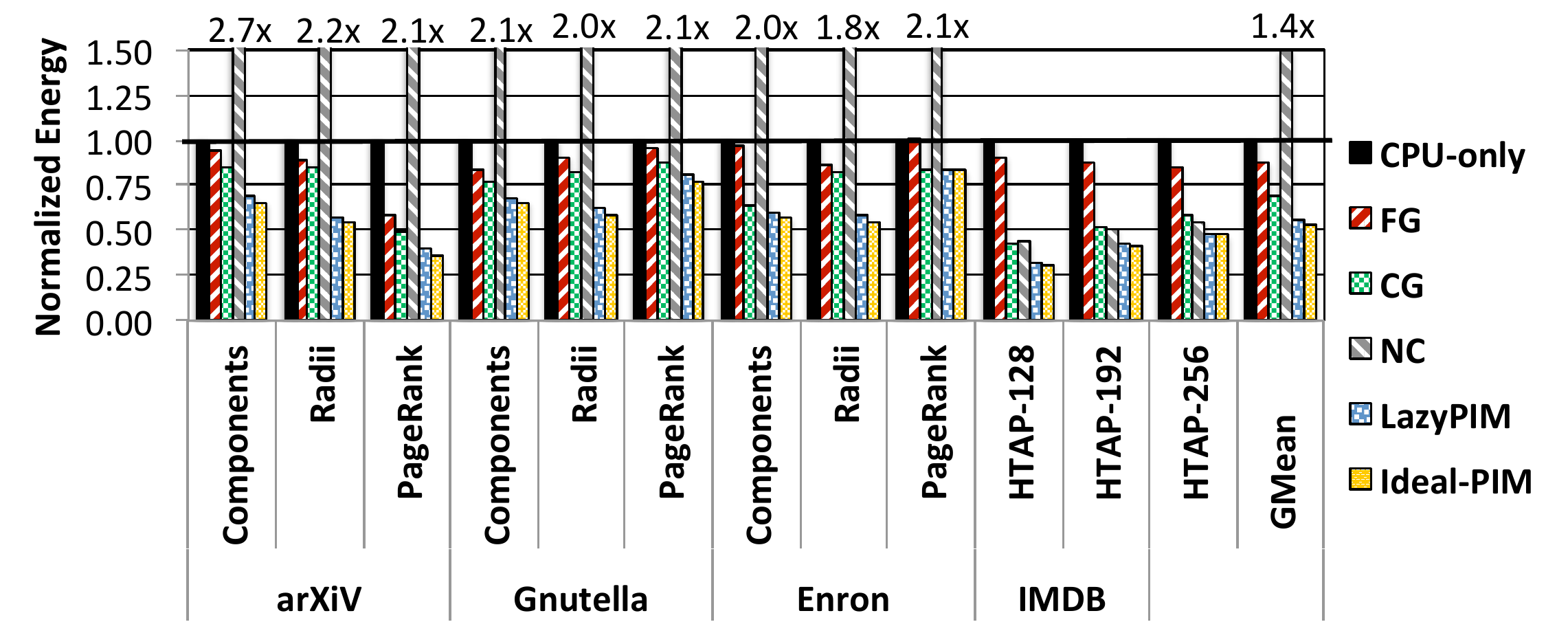}%
    \vspace{-8pt}%
    \caption{Energy for all applications with 16~threads normalized to CPU-only execution.}
    \label{fig:energy}
    \vspace{-5pt}
\end{figure}

Figure~\ref{fig:energy} shows the energy consumption of the PIM coherence 
approaches for our 16-core system, normalized to CPU-only, across all of
our applications and input sets. \changes{We find that LazyPIM comes within 4.4\% of Ideal-PIM,
and significantly reduces
energy consumption, by an average of of 35.5\% over FG, 18.0\% over CG,
62.2\% over NC, and 43.7\% over CPU-only.}  The main energy improvement for
LazyPIM over CPU-only comes from a reduction in interconnect and memory
utilization, as LazyPIM successfully reduces data movement across the off-chip
channel.

Other PIM coherence approaches are unable to retain these benefits, as FG, CG, 
and NC all consume more energy due to their poor handling of coherence.
CG has a large number of writebacks, which leads to increased L1 and L2 cache
lookups and a large amount of traffic sending dirty data back to DRAM 
preemptively.  Thanks to these writebacks, the energy consumption of the HMC
memory alone increases by 18.9\% on average over CPU-only, canceling out CG's
net savings in interconnect energy (which was reduced by 49.1\% over CPU-only).
The energy consumption of NC suffers greatly from the large number of memory
operations that are redirected from the processor caches to DRAM.  While the 
cache energy consumption falls by only 7.1\%, interconnect and HMC energy
increase under NC by 3.1x and 4.5x, respectively.
For FG, the large number of coherence messages transmitted between the 
processor and the PIM cores undoes \changes{a significant amount of} the energy benefits of moving computation
to DRAM.

We conclude that the speculative coherence update approach used by LazyPIM is
effective at retaining the energy benefits of PIM execution.

\subsection{Effect of Partial Kernel Commits}
\label{sec:eval:partial}

\begin{figure}[t]
    \centering
    \includegraphics[width=0.9\linewidth]{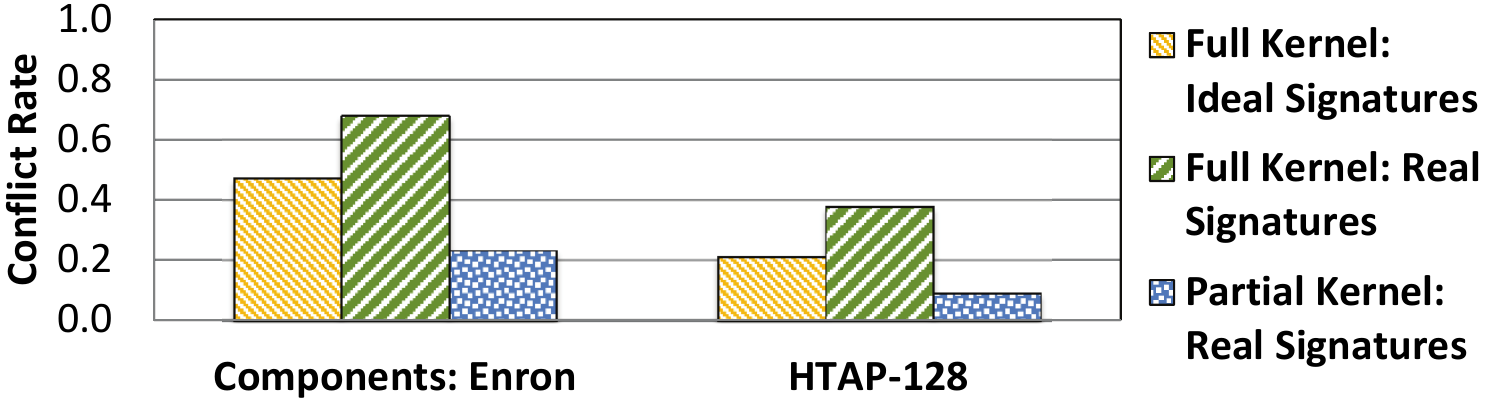}
    \vspace{-5pt}%
    \caption{Effect of full kernel commits vs.\ partial kernel commits on conflict rate of representative 16-thread workloads.}
    \label{fig:conflictrate}
    \vspace{-9pt}
\end{figure}

As we discussed in Section~\ref{sec:mech:partial}, dividing a PIM kernel into
smaller partial kernels allows LazyPIM to lower both the probability of conflicts and false positives. 
Figure~\ref{fig:conflictrate} shows
the extent of the benefits that partial commits offer,   
for a signature size of 2~Kbits, for two representative 16-thread workloads:
Components using the Enron graph, and HTAP-128.  
If we study an idealized version of full kernel commit, where no false 
positives exist, we find that a relatively high percentage of commits contain
conflicts (47.1\% for Components and 21.3\% for HTAP).  Using realistic
signatures for full kernel commit, which includes the impact of false 
positives, the conflict rate increases to 67.8\% for Components and 37.8\% for
HTAP.

Whenever a conflict is detected, a rollback must be initiated.  In the case of 
the full kernel signatures above, this means that approximately half of all
commit attempts force LazyPIM to roll back to the beginning of the entire
kernel.  
As Figure~\ref{fig:conflictrate} shows, our partial kernel commit
technique significantly reduces this burden. Even factoring in false 
positives, the conflict rate drops to 23.2\% for Components and 9.0\% for
HTAP.  Further helping partial kernel commit is the fact that in the cases
when rollbacks do occur, the length of the rollback is smaller, as only a
small portion of the kernel must be re-executed. We conclude that partial kernel commit is an effective technique at keeping
the overheads of LazyPIM rollback low.

\subsection{Effect of Signature Size}
\label{sec:eval:signature}

\begin{figure}[h]
    \vspace{-3pt}
    \centering
    \includegraphics[width=0.85\linewidth]{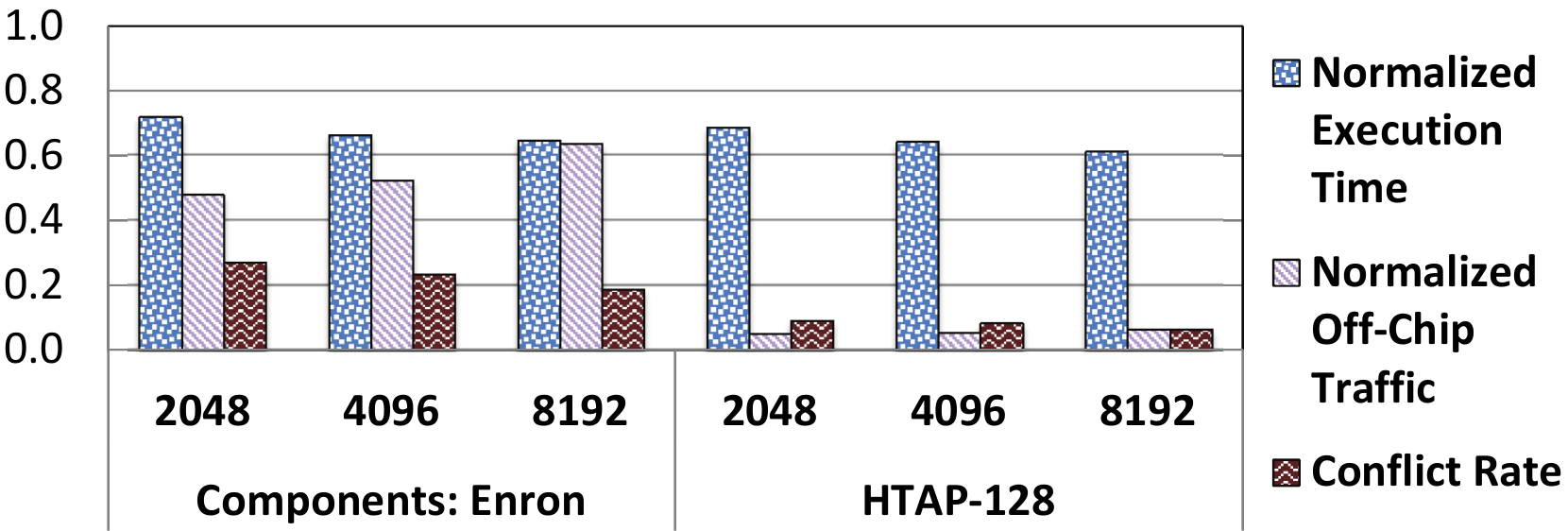}
    \vspace{-6pt}%
    \caption{Effect of signature size on representative 16-thread workloads, with execution time and off-chip traffic normalized to CPU-only execution.}
    \label{fig:sigsize}
    \vspace{-3pt}
\end{figure}

We study the impact of signature size on \changes{execution time, 
off-chip traffic, and the conflict rate, using two representative 16-thread workloads:
Components using the Enron graph, and HTAP-128.}  Note that for all signature sizes, we
continue to store 250~addresses per signature using partial kernel commit.  
Figure~\ref{fig:sigsize}
shows these three factors, with execution time and off-chip traffic normalized
to CPU-only, for our two representative 16-thread workloads. 
Increasing the signature size from 2~Kbits to 8~Kbits reduces the
conflict rate by 30.0\% for Components and by 29.3\% for HTAP, which in turn
results in a reduction in execution time of 10.1\% and 10.9\%, respectively,
as the lower conflict rate results in a smaller number of rollbacks.  The
reduction in conflict rate is a direct result of the lower false positive rate
with 8~Kbit signatures (a reduction of 31.4\% for Components and 40.5\% for
HTAP).  However, this comes at the cost of increased off-chip 
traffic, as the 8~Kbit signature requires 32.7\% more traffic for Components
and 31.4\% more traffic for HTAP.

%

From our evaluations, we conclude that using a smaller 2~Kbit signature strikes
a good balance between execution time and off-chip traffic across all of our
workloads.


\section{Related Work}
\label{sec:related}

\changes{Most recent PIM proposals assume that there is only a limited amount
of data sharing between PIM kernels and the processor threads of an
application. As a result, they employ a variety of na\"ive solutions to
maintain coherence, such as (1)~assuming that PIM kernels and processor threads
never access data concurrently~\cite{PICA,Ahn:2015:SPA:2749469.2750386},
(2)~making PIM data non-cacheable within the
processor~\cite{Ahn:2015:SPA:2749469.2750386,7056040,pugsley2014ndc},
(3)~flushing \emph{all} dirty cache lines in the processor before starting PIM
execution~\cite{toppim,Mingyu:PACT,Lee2001AutoMapping}, or (4)~locking
\emph{all} of the PIM data during PIM execution~\cite{7056040,JAFAR}.
We already showed in Section~\ref{sec:motivation} that the non-cacheable (2) and coarse-grained approaches (3,4)
to coherence do not work well within the context of PIM.

Other proposals in related domains attempt to efficiently support coherence.
HSC~\cite{HSC} reduces coherence traffic between the CPU and the GPU, but the
trade-offs in the HSC CPU-GPU system are fundamentally different than in a
CPU-PIM system. Unlike in a CPU-PIM system, which requires off-chip
communication, both the CPU and the GPU in HSC are on-chip and are coherent
through the last-level cache, making communication much cheaper.
More importantly, HSC employs coarse-grained coherence to reduce the coherence
traffic from the GPU to the directory, which we have demonstrated is unable to
retain the benefits of PIM. 
FUSION~\cite{FUSION} mitigates data movement between the CPU and on-chip
accelerators and decreases the communication traffic between 
accelerators. FUSION employs traditional MESI (FG) coherence for communication
between accelerators and the CPU. Because the accelerators are \emph{on-chip},
the cost of data movement between the CPU and the accelerators is roughly the
same as that between the accelerators themselves. However, in a CPU-PIM system,
the cost of communication between the processor and the PIM cores over the
\emph{off-chip} channel far outweighs the cost of communication between PIM
cores, and this off-chip communication cost can significantly degrade the
benefits of using PIM. We showed in Section~\ref{sec:motivation} that adopting FG
coherence for a CPU-PIM system eliminates most benefits of PIM.

Other works~\cite{rsp, denovo, hlrc} focus on reducing on-chip intra-GPU
communication and coherence traffic. These works rely heavily on software
assistance to reduce coherence complexity, which requires considerable
programmer and compiler involvement (e.g., custom APIs, custom programming
models), preventing their applicability to all types of applications. In
contrast, LazyPIM can enable efficient off-chip coherence for \emph{any} type
of application by employing speculation coupled with simple programmer
annotation. These works all attempt to reduce on-chip coherence, which is
orthogonal to the goals of LazyPIM. While these works can be applied in
conjunction with LazyPIM to reduce inter-PIM coherence traffic, they are not
effective at reducing the off-chip coherence traffic that LazyPIM works to
minimize.}


\section{Conclusion}
\label{sec:conclusion}

We propose LazyPIM, a new cache coherence mechanism  
for PIM architectures.  Prior approaches to PIM
coherence generate very high off-chip traffic for important
data-intensive applications.  LazyPIM avoids this by avoiding coherence lookups
\emph{during} PIM kernel execution.
The key idea is to use compressed coherence \emph{signatures} to batch the lookups and verify correctness \emph{after} 
the kernel completes. 
As a result of the more efficient approach to coherence employed by LazyPIM,
applications that performed poorly under prior approaches to PIM coherence
can now take advantage of the benefits of PIM execution.
LazyPIM improves average performance by 19.6\%
reduces off-chip traffic by 30.9\%, and reduces energy consumption
by 18.0\% over the best prior approachs to PIM coherence, 
while retaining the conventional multithreaded programming model.

\vspace{-2pt}


\newpage
\bibliographystyle{IEEEtranS}
\bibliography{references}

\end{document}